\documentclass[fleqn,usenatbib]{mnras}
\usepackage{newtxtext,newtxmath}
\usepackage[T1]{fontenc}
\DeclareRobustCommand{\VAN}[3]{#2}
\let\VANthebibliography\thebibliography
\def\thebibliography{\DeclareRobustCommand{\VAN}[3]{##3}\VANthebibliography}

\usepackage{graphicx}	
\usepackage{graphics}
\usepackage{amsmath}	
\usepackage{lipsum}
\usepackage{caption}
\usepackage{listings}
\usepackage{comment}
\usepackage{multirow}
\usepackage{xcolor}
\usepackage{subcaption}
\usepackage{mwe}
\graphicspath{{./Pictures/}}
\pdfoutput=1
\usepackage{soul}

\newcommand{\tvec}[1]{\mathbf{#1}} 

\newcommand{\de}{\mathrm{d}}

\def\mean#1{\left< #1 \right>} 

\newcommand{\Omm}{\Omega_\mathrm m}
\newcommand{\Omb}{\Omega_\mathrm b}
\newcommand{\ASN}[1]{A_{\mathrm{SN}#1}}
\newcommand{\AAGN}[1]{A_{\mathrm{AGN}#1}}

\newcommand{\Ill}{IllustrisTNG }

\newcommand{\IllCV}{IllustrisTNG-CV }
\newcommand{\IllLH}{IllustrisTNG-LH }
\newcommand{\IllP}{IllustrisTNG-1P }
\newcommand{\SIMCV}{SIMBA-CV }
\newcommand{\IllBig}{IllustrisTNG-300}

\newcommand{\logNS}{\log N-\log S_B}
\newcommand{\SB}{S_B}
\newcommand{\Rvir}{R_\mathrm{200m}}
\newcommand{\Rsp}{R_\mathrm{sp}}
\newcommand{\unitsSB}{\mathrm{ph \ cm^{-2} \ s^{-1} \ sr^{-1}}}

\title[WHIM in emission]{Studying the Warm Hot Intergalactic Medium in emission: a reprise}

\author[G. Parimbelli et al.]{
G. Parimbelli$^{1,2,3,4,5,6}$ \thanks{E-mail:gabriele.parimbelli@edu.unige.it},
E. Branchini$^{1,2,3}$,
M. Viel$^{4,7,6,5}$,
F. Villaescusa-Navarro$^{8,9}$,
J. ZuHone$^{10}$
\\
$^{1}$ Dipartimento di Fisica, Universit\`{a} degli studi di Genova, via Dodecaneso 33, I-16146, Genova, Italy\\
$^{2}$ Dipartimento di Matematica e Fisica. Universit\`{a} Roma Tre, via della Vasca Navale 84, I–00146, Roma, Italy\\
$^{3}$ INFN – Istituto Nazionale di Fisica Nucleare, via della Vasca Navale 84, I–00146 Roma, Italy\\
$^{4}$ Scuola Internazionale Superiore di Studi Avanzati. via Bonomea, 265, I–34136, Trieste, Italy\\
$^{5}$ INAF-OATs, Osservatorio Astronomico di Trieste, via Tiepolo 11, I–34131 Trieste, Italy\\
$^{6}$ IFPU – Institute for Fundamental Physics of the Universe, via Beirut 2, I–34151 Trieste, Italy\\
$^{7}$ INFN – Istituto Nazionale di Fisica Nucleare, via Valerio 2, I-34127 Trieste, Italy\\
$^{8}$ Center for Computational Astrophysics, Flatiron Institute, 162 5th Avenue, New York, NY, 10010, USA\\
$^{9}$ Department of Astrophysical Sciences, Princeton University, Peyton Hall, Princeton NJ 08544, USA\\
$^{10}$ Center for Astrophysics, Harvard \& Smithsonian, 60 Garden Street, Cambridge, MA 02138, USA\\
}

\date{Accepted XXX. Received YYY; in original form ZZZ}

\pubyear{2022}

\begin{document}
\label{firstpage}
\pagerange{\pageref{firstpage}--\pageref{lastpage}}
\maketitle

\begin{abstract}
The  Warm-Hot Intergalactic Medium (WHIM) is believed to host a significant fraction of the ``missing baryons'' in the nearby Universe.
Its signature has been detected in the X-ray absorption spectra of distant quasars. 
However, its detection in emission, that would allow us to study the WHIM in a systematic way, is still lacking.
Motivated by the possibility to perform these studies with next generation integral field spectrometers, and thanks to the availability of a large suite of state-of-the-art hydrodynamic simulations -- the CAMELS suite -- we study here in detail the emission properties of the WHIM and the possibility to infer its physical properties with upcoming X-ray missions like Athena.
We focused on the two most prominent WHIM emission lines, the OVII triplet and the OVIII singlet, and build line surface brightness maps in a lightcone, mimicking a data cube
generated through integral field spectroscopy.
We confirm that detectable WHIM emission, even with next generation instruments, is largely associated to galaxy-size dark matter halos and that the WHIM properties evolve little from $z\simeq0.5$ to now.
Some characteristics of the WHIM, like the line number counts as a function of their brightness, depend on the specific hydrodynamic simulation used, while others, like the WHIM clustering properties, are robust to this aspect.
The large number of simulations available in the CAMELS datasets allows us to assess the sensitivity of the WHIM properties to the background cosmology and to the energy feedback mechanisms regulated by AGN and stellar activity.
Interestingly, the number of WHIM line counts, especially the bright ones, is significantly affected by the details of the feedback mechanisms while the spatial clustering properties are not. The opposite occurs with the cosmological parameters: they affect the spatial distribution of the WHIM but not the counts.
As a result, emission studies have the potential to separate the cosmological aspects from the baryonic processes and set constraints on the latter.
Finally, we provide some reasonable, though not fully realistic, forecast on
WHIM observations with a typical next generation spectrometer like X-IFU on board of Athena.
We expect to detect, with a significance larger than 3-$\sigma$, about 1--3 WHIM lines in emission per pixel, independently on the angular resolution of the instrument. Furthermore, we expect to measure the angular correlation of the WHIM signal and to trace the emission profile of the WHIM around halos out to a few tens of arcmin, well beyond the size of a typical WHIM emitter.
\end{abstract}




\section{Introduction}
\label{sec:introduction}

The observed abundance, in the spectra of distant quasars,  of the light elements produced during the big bang nucleosynthesis has provided us with a spectacular and precise estimate of the comoving abundance of baryonic matter in the Universe \citep{BBN_review} at that epoch.
A measurement that is in very good agreement with the estimate of the baryon abundance at the epoch of the matter-radiation decoupling obtained by analyzing the angular spectrum of the temperature fluctuations in the cosmic microwave background
\citep{Planck2018}.
The discovery that, instead, about half of the expected baryons are missing from the observed total mass budget in the local Universe came as a surprise and constitutes what has been dubbed the  ``Missing Baryons'' (MB) problem ever since \citep{missing_baryons-Fukugita+98,missing_baryons-Cen+99}.
More recent observations confirm that the problem is still there and that, despite dedicated searches, about 30\% of the baryons in the local Universe still evade detection \citep{missing_baryons-Shull+12}.

Hydrodynamic simulations designed to follow the evolution of cosmic structures in large regions of a  $\Lambda$CDM Universe \citep{missing_baryons-Cen+99}
have indicated the solution to the problem:  
a sizable fraction of baryons in the low redshift Universe is in a gaseous form and traces the large scale filaments and the external regions of the dense knots that constitute the cosmic web. 
This gas is shock-heated and ionized during the buildup of the large scale cosmic structures.
As a result, most baryons are now expected to be in the form of a highly ionized, warm-hot 
diffuse gas with temperatures in the range $T \sim 10^5-10^7$ K and number densities of $n \sim 10^{-6}-10^{-4}$ cm$^{-3}$: the ``warm-hot intergalactic medium'' (WHIM).
Because of its physical state, it is very hard to detect, since its high ionization state leaves no signature in the local Lyman-$\alpha$ forest and because of its low density its bremsstrahlung emission is to faint to be detected in the X-ray continuum.
One of the best strategies to detect the missing baryons is, instead, to observe the characteristic absorption or emission lines in the X-ray band
of the highly ionized ``metals'' produced in the stars and then diffused in the intergalactic medium by galactic winds and supernovae explosions.

Since detecting absorption lines in the spectra of bright background X-ray sources is comparatively easier than observing the WHIM lines in emission, most of the theoretical studies (e.g. 
\cite{Cen+06, Branchini+09,Tepper-Garcia+11,Cen+12,missing_baryons-Shull+12,Rahmati+16,Oppenheimer+16,Oppenheimer+18,Nelson+18,martizzi19,wijers19})
and observational efforts  (see e.g. \cite{Nicastro2017} and references therein)
have been focusing on this possibility.
This strategy proved to be successful and eventually led to the unambiguous detections of the WHIM in the X-ray spectra of two bright quasars \citep{WHIM_detection-Nicastro+18,WHIM_detection-Kovacs+19}.

These important discoveries have validated the WHIM model that has been built from the hydrodynamic simulations and, more in general, have confirmed the correctness of the cosmological structure formation model in the $\Lambda$CDM framework.
However, the paucity of the WHIM line detections and the very nature of the absorption studies that only allows one to probe the WHIM along a few, pre-selected directions, makes it difficult to probe the physical status and spatial distributions of a large fraction of the cosmic baryons in the low-redshift Universe.

Studying the WHIM in emission could, in principle, significantly improve our knowledge of the missing baryons physics for several reasons.
First of all, the use of integral field spectroscopy allows one to probe the distribution of the missing baryons over any volume, and not just along the sightlines to bright sources. The technique is similar to that of the line emission mapping with the 21 cm line in the radio band, except that in this case one should use the metal lines in the soft X-ray band. 
Furthermore, absorption and emission line studies are highly complementary. Absorption is proportional to the local density of the gas and preferentially occurs in in regions of moderate overdensity. On the contrary, the emission signal is approximately proportional to the square of the density and, therefore, preferentially produced in high density environments, more sensitive to stellar feedback and AGN activity. 

What hampered emission studies so far is the difficulty to disentangle the WHIM signal from foreground emission, which is typically dominant. As a result, only a few, somewhat controversial, claims of WHIM detection in emission have been made
(i.e. \cite{WHIM_detection-Nicastro+05,WHIM_detection-Zappacosta+05,WHIM_detection-Werner+08,bonamente22}) and turned out to be produced by gas
with density and temperature higher than the typical WHIM and associated to the external parts of virialized objects, like galaxy clusters.
As a result, a comparatively smaller number of studies have been dedicated to modeling the WHIM in emission using hydrodynamic simulations \citep{Cen+06,Metal_line_emission-Bertone+09,WHIM-Takei+11,Ursino+11,Cen+12,Roncarelli+12,Nelson+18} and even less have contemplated the possibility to combine absorption and emission studies (e.g. \cite{Branchini+09}).
These studies are also tightly related to the circumgalactic medium, whose presence can be detected both in emission \citep{CGM_EAGLE_emission-Wijers+22} and absorption \citep{CGM_EAGLE_absorption-Wijers+20}.

The approaching of the new era of integral field spectroscopy in the X-ray band with the planned XRISM and Athena space missions \citep{Xrism+20,Athena}, as well as the Line Emission Mapper (LEM) Probe concept\footnote{\url{https://lem.physics.wisc.edu}}, has triggered a renewed interest in the subject that justifies a fresher and deeper look into the subject.
A significant update of the previous studies, which constitutes the main goal of this paper, is made possible thanks to the availability of a new, publicly available, large set of state-of-the-art hydrodynamic simulations, the CAMELS suite \citep{CAMELS_simulations,CAMELS_simulations_data_release}. These simulations allow us to overcome one potential issue of the previous analysis: the fact that predictions of the WHIM emission properties were based on a single or a handful of numerical experiments, while we know that theoretical uncertainty significantly contribute, and possibly dominate, the error budget.
Thanks to the CAMELS suite in this work we will be able to span the large space of the parameters that drive the evolution of the mass density field and that regulate the energy feedback processes. 
Moreover, thanks to the presence of two different sets of simulations performed with two different hydrodynamic codes, AREPO \citep{Arepo_public} and GIZMO \citep{Hopkins2015_Gizmo}, we will be able to appreciate the sensitivity to the numerical technique used to simulate the WHIM.

We aim at understanding the properties of the WHIM in emission through the common summary statistics that will be computed from the X-ray surface brightness maps generated by integral field spectroscopy, namely the number counts of the characteristics WHIM emission line (and for the sake of simplicity we will focus on the most prominent ones: 
on the OVII line triplet with rest frame energies $E_\mathrm{OVII} = 0.56098, 0.56874, 0.57395$ keV  and the OVIII singlet at $E_\mathrm{OVIII} = 0.65355$ keV) and the angular 2-point correlation function of these lines and their cross-correlation with galaxies (and their halo host). 
For this we will need to specify the characteristics of the detector. We will assume that observations will be performed with a calorimeter spectrometer similar to X-IFU on board of the Athena satellite. We stress, however, that our goal here is not to provide a realistic forecast for future X-ray analyses but only to obtain reasonable estimates on the precision with which summary statistics will be measured and our ability to use them to characterize the physics of the missing baryons.

This paper is organized as follows.
In section \ref{sec:dataset} we present the subset simulations that we have selected from
the CAMELS suite and that we use in our analysis, discussing the criteria for this choice.
In section \ref{sec:models} we describe the procedure adopted to generate the simulated X-ray emission maps from the hydrodynamic simulation outputs. Some simplifying assumptions have been made in the process that we discuss here. In this section we also describe the estimator used to compute the summary statistics considered in this work.
In section \ref{sec:results} we present the results of our analysis and test their robustness 
to various physical and numerical effects.
In section \ref{subsec:lightcone} we provide some forecast on the WHIM line detectability and compare our results with those of \cite{WHIM-Takei+11}.
Finally, in section \ref{sec:conclusions} we discuss our results and draw our conclusions.

\section{Data sets}
\label{sec:dataset}

In this section we describe the simulated datasets that we use in this work to generate dark matter halo catalogs, mock WHIM spectra, and surface brightness maps.


\subsection{The CAMELS Parent Dataset}
\label{subsec:simulations}

\begin{table*}
	\centering
	\begin{tabular}{ccccccc}
		\hline
		    \textbf{Name} & \textbf{N. realizations} & \textbf{random seeds} &
		    \textbf{$\Omm$} & \textbf{$\sigma_8$} & \textbf{$\ASN{1}$, $\AAGN{1}$} & \textbf{$\ASN{2}$, $\AAGN{2}$} \\
		\hline 
		\hline
		    \textbf{\IllCV} & 27 & Several & 0.3 & 0.8 & 1 & 1  \\
	    \hline 
	        \textbf{\SIMCV} & 27 & Several & 0.3 & 0.8 & 1 & 1 \\
	    \hline
	        \textbf{\IllLH} & 63 & Several & [0.25, 0.35] & [0.75, 0.85] & [0.25, 4] & [0.5, 2] \\
	    \hline
	        \textbf{\IllP}  & 45 & Same & [0.25, 0.35] & [0.75, 0.85] & [0.25, 4] & [0.5, 2] \\
	    \hline
	        \textbf{\IllBig} & 1 &  & 0.3089 & 0.8159 & 1 & 1 \\
	    \hline
	\end{tabular}
	\caption{Subsets of CAMELS simulations used in this work. Column 1: Set name. Column 2: Number of simulations. Column 3: Random seed(s) used to set the initial conditions. Column 4: $\sigma_8$ range of values.  Column 5: $\Omm$ range of values.  Column 6: $\ASN{1}$ and $\AAGN{1}$ range of values. The two parameters are set equal. Column 7: $\ASN{2}$ and $\AAGN{2}$ range of values. The two parameters are set equal. 
}
	\label{tab:simulation_sets_table}
\end{table*}

The parent dataset is the CAMELS suite of cosmological simulations \citep{CAMELS_simulations} which consists of  4,233 simulations, about half of which are dark-matter-only and half (magneto)-hydrodynamic, designed to train machine learning techniques and applications to astrophysical datasets.
The simulations follow the evolution of cosmic structures from $z=127$ to present day using 256$^3$ dark matter (DM) particles as well as, when present, 256$^3$ fluid particles in a cubic box of 25 comoving Mpc/$h$.
All simulations assume a $\Lambda$CDM model.
For each realization, 34 snapshots are created at various redshifts. Since we are interested in studying the WHIM which is a relatively local phenomenon, we only use the ten snapshots at $z \leq 0.54$ whose characteristics are specified in Table~\ref{tab:snapshot_grid}.
To avoid divergence when creating the flux maps used in this work we place the $z=0$ snapshot at a distance of $\sim 118 $ Mpc/$h$ (the precise value depending on the cosmology assumed) from the observer, corresponding to  $z=0.04$. 

The suite is designed to span a wide region of a 6-dimensional parameter space.
Two dimensions are meant to explore different cosmological models, characterized by the mean matter density $\Omm$ and the clustering amplitude $\sigma_8$. The remaining 4 dimensions are meant to sample different energy feedback models.
Two of them regulate the stellar feedback from supernovae (parameter $\ASN{1}$) and from galactic wind ($\ASN{2}$).
The other two parametrize AGN feedback through energy-momentum injection ($\AAGN{1}$) and  and jet speed ($\AAGN{2}$). 
These parameters measure the relative variations with respect to the reference feedback models used in the IllustrisTNG and SIMBA simulations \citep{IllustrisTNG_simulations,SIMBA_simulations,CAMELS_simulations}.
Their values are normalized accordingly.

In this work we also use catalogs of DM halos with masses in the range $\sim 4 \times 10^8-5\times 10^{13} \ M_\odot/h$ that have been identified using the friends-of-friends algorithm \citep{FoF-Huchra+82,FoF-Davis+85} and the SUBFIND software \citep{SUBFIND-Springel+01}.

All cosmological parameters but $\sigma_8$ and $\Omm$ are kept fixed in the analysis. They are: 
$\Omb = 0.049$, $h=0.6711$, $n_s = 0.9624$.
The value of $\Omb$ implies an initial mass of $1.27 \times 10^7 \ M_\odot/h$ for the gas particles.

The CAMELS suite includes two types of hydrodynamic simulations, that we use in this work:
\begin{itemize}
    \item \textbf{IllustrisTNG}: These are 1,092 hydrodynamic simulations run with the AREPO code \citep{Arepo,Arepo_public} that implements the same galaxy formation model and subgrid physics as the original \Ill simulations of \cite{IllustrisTNG_simulations,Weinberger+17,IllustrisTNG_stellar-Pillepich+18} 

    \item \textbf{SIMBA}: This is a matching set of 1,092 hydrodynamic simulations run with the GIZMO code \citep{Hopkins2015_Gizmo} that implements the same galaxy formation model and subgrid physics as the original SIMBA simulations of  \cite{SIMBA_simulations}.
\end{itemize}

\subsection{Datasets used in this work}
\label{subsec:subsets}

Each dataset described above is a suite of simulations from which we selected a few for our purposes. The subset of simulations used in this work consists of the following samples:
\begin{itemize}
    \item \textbf{\IllCV} and \textbf{\SIMCV} samples. These are 27 + 27 simulations extracted from both the \textbf{IllustrisTNG} and the \textbf{SIMBA} suites.
    All of them use the same set of cosmological and astrophysical parameters but different random seeds to set initial conditions.
    The relevant parameters that characterize these samples, which we regard as the set of fiducial parameters, are listed in Table~\ref{tab:simulation_sets_table}. 
    The results obtained from this samples will allow us to evaluate the effect of the cosmic variance and to estimate its contribution to the total error budget.
    \item \textbf{\IllLH}. This set consists of 1,000 realizations selected from the \textbf{IllustrisTNG} sample. 
    These simulations have been evolved from different initial conditions and by using different values for the 6 free parameters (of both cosmological and astrophysical type) chosen within a pre-defined range around the fiducial values.
    The resulting 6-dimensional parameter space is very wide, since it was designed for machine learning purposes.
    However, in this work we are interested in studying the properties of the WHIM in a realistic cosmological setup that matches observations. Therefore, we decided to explore a limited region of the parameter space in which {\it i)} the value of $\Omm$ is consistent with the results of the 
    3x2pt joint weak lensing and clustering analyses of the DES-year 3 catalog \citep{DES_yr3} and  {\it ii)} the value of $\sigma_8$ is consistent with those obtained by the Planck Team  \citep{Planck2018}.
    As a result we consider $\Omm$ and $\sigma_8$ values in the ranges [0.25,0.35] and [0.75,0.85], respectively.
    Once these selections criteria are applied we are left with 63 simulations that we use to evaluate the uncertainties in predicting the WHIM properties, including uncertainties in modeling stellar and AGN feedback mechanisms. 
    \item \textbf{\IllP }: The simulations in this subset have been performed varying one parameter at the time and using the same initial conditions.
    11 different values are used for each of the 6 parameters, totaling to 61 simulations.
    We consider only the parameters whose values are in the same range as in the \textbf{\IllLH} sample, for a total 45 simulations.
\end{itemize}

In this work we consider an additional dataset which is not part of the CAMELS suite: the \textbf{\IllBig} hydrodynamic simulation \citep{IllustrisTNG_simulations,IllustrisTNG_stellar-Pillepich+18,IllustrisTNG_colour_bimodality-Nelson+18,IllustrisTNG_matter-Springel+18,IllustrisTNG_Mg_Eu-Naiman+18,IllustrisTNG_magnetic_field-Marinacci+18}.
This is a single numerical experiment performed in a  box of side 205 comoving Mpc/$h$, i.e. $\sim \ 8$ times larger than the CAMELS one, that assumes a \cite{Planck2016} flat $\Lambda$CDM cosmology with $\Omm=0.3089,\Omb=0.0486,h=0.6774,n_\mathrm s=0.9667, \sigma_8=0.8159$. 
We use this simulation to assess the effects of the large wavelength modes, that are missing from the CAMELS dataset, and of the mass resolution, which is higher in this case because the number of particles in this simulation is about 1000 larger than in the CAMELS
(the mass of a single gas particle is $7.6\times 10^6 \ M_\odot/h$).
Moreover, performing the analysis in a volume $\sim 550$ times larger than that of the CAMELS will allow us to better samples rare events associated to very massive halos and, in general, to improve statistics.


\section{Modeling the WHIM properties}
\label{sec:models}

We now use the hydrodynamic simulations to model the OVII and OVIII line emission maps (section~\ref{subsec:line_SB}) and to compute  summary statistics such as emission line counts and 2-point angular correlation (section~\ref{subsec:statistics}).

\subsection{Line surface brightness maps}
\label{subsec:line_SB}

The WHIM X-ray spectrum is characterized by the emission (or absorption) lines of several highly ionized ``metals'' such as carbon, oxygen, neon, iron and magnesium.
Here we will focus on the most prominent line systems: the OVII triplet at 0.561, 0.569, 0.574 keV and the OVIII singlet at 0.653 keV.
To model the X-ray emission spectra from the simulation outputs we proceed as follows:
\begin{itemize}
    \item We divide the gas particles in the simulations in two subsets: the WHIM and non WHIM particles. WHIM particles have gas overdensity $\delta<1,000$ and temperature $10^5 \ \mathrm{K} < T < 10^7 \ \mathrm{K}$. All remaining particles constitute the non-WHIM gas. The definition is somewhat arbitrary but matches that used in several works in the literature.
    \item For each gas particle, WHIM and non-WHIM, we
    compute the emissivity spectra by feeding density, temperature and metallicity of the gas particle to the \texttt{pyXSIM} code \citep{pyxsim-ZuHone+16}.
    To express the absolute metallicity of the gas into solar units we used the tables in \cite{Asplund_tables+09}.
    In the \texttt{pyXSIM} code the OVII and OVIII ion fractions are computed assuming collisional ionization equilibrium using the \texttt{AtomDB} tables and \texttt{APEC} emission model \citep{Smith2001,Foster2012}, i.e. we ignore  photo-ionization.
    This assumption, which has been adopted in most of the previous emission studies of the WHIM \citep{WHIM-Takei+11,Roncarelli+12}, is satisfied at low redshifts for hydrogen gas density $n_{H}>10^{-4}$ (see e.g. Fig. 3 of \cite{Nelson+18}) and therefore fully justified in the high-density and high-temperature environments where the oxygen line emission is strongest and  potentially detectable by future observations.
    \item Once the emission spectrum for each particle is computed, we estimate the OVII and OVIII line emissivity $j(\rho,T,Z)$ by estimating the energy contributed by a portion of the spectrum centered around the emission line and slightly larger than the width of the line.
    \item We then compute the photon flux of the oxygen lines from each gas element as  \citep{Metal_line_emission-Bertone+09}:
    \begin{equation}
        \Phi_\mathrm{ph} = \frac{j(\rho,T,Z)}{4\pi D^2_L(z)} \ \frac{m}{\rho} \ \frac{1+z}{\mean{E}} \, ,
        \label{eq:photon_flux}
    \end{equation}
    where $m$ is the mass of the gas particle at the observed redshift $z$ and luminosity distance $D_L(z)$, $\rho$ is the gas density and $\mean{E}$ is the mean rest frame energy in the spectral range considered.
    \item The OVII and OVIII line fluxes of all particles are then interpolated on a grid, summed over and divided by the solid angle subtended by the pixel to obtain the OVII and OVIII surface brightness maps. 
\end{itemize}

\begin{table}
\centering
\resizebox{\columnwidth}{!}{
\begin{tabular}{||c | c c c c ||}
 \hline
    \multirow{2}{*}{$\mathbf{z}$} & 
    \textbf{comoving distance} &
    \textbf{grid size} &
    \textbf{slice thickness} &
    \textbf{field of view}\\
    & $D_L(z)$ [Mpc/$h$] & $N_\mathrm{grid}$ & $\Delta L$ [Mpc/$h$] & FoV [deg$^2$]\\
 \hline
 0.54 & 2163.0 &   732 & 14.4 &    1.0 \\ 
 0.47 & 1822.8 &   829 & 14.3 &    1.3 \\
 0.40 & 1508.9 &   955 & 14.2 &    1.8 \\
 0.33 & 1223.1 &  1123 & 14.0 &    2.4 \\
 0.27 &  963.0 &  1360 & 13.8 &    3.6 \\
 0.21 &  727.3 &  1716 & 13.6 &    5.7 \\
 0.15 &  514.6 &  2313 & 13.4 &   10.3 \\
 0.10 &  323.3 &  3509 & 13.1 &   23.8 \\
 0.05 &  152.3 &  7104 & 12.8 &   97.4 \\
 0.04 &  123.6 &  8679 & 12.8 &  145.3 \\ [1ex] 
 \hline
\end{tabular}
}
\caption{The table describes the specifics of the 10 snapshots we consider in our analysis. The first column reports the redshifts of the snapshots. The other columns report the comoving distance to the redshift, the size of the grid employed to build the maps, the thickness of the slice corresponding to the spatial resolution along the line-of-sight and the field of view that the simulation represents, respectively.
All these quantities have been computed using the fiducial cosmology of the CV subsets.}
\label{tab:snapshot_grid}
\end{table}

\begin{table}
\centering
\begin{tabular}{||c c||} 
 \hline
 \multicolumn{2}{||c||}{\textbf{X-IFU specifics}} \\
 \hline
 Angular resolution  & $\theta_\mathrm{res}= 5''$ \\
 Energy resolution   & $\Delta E= 2.5$ eV \\
 Detection threshold & \multirow{2}{*}{$S_\mathrm{B,min} = 0.1 \ \unitsSB$} \\
 ($t_\mathrm{obs} = 100$ ks) & \\
 Field of view   & (5 arcmin)$^2$ \\
 \hline
\end{tabular}
\caption{Specific of the Athena X-IFU instrument \citep{Athena}.}
\label{tab:instrument_specific}
\end{table}

The final map consists of a 3D data cube in which each pixel specified the OVII and OVIII surface brightness at a given angular position and contributed by the gas at a given redshift.
Similar 3D data cubes will be generated by calorimeters on board of next generation X-ray satellite, designed to perform spatially resolved spectroscopy. 
Therefore, to add a realistic touch to our results, we decided to set the size of the 3D pixels by matching the expected angular resolution $\theta_\mathrm{res}$ and energy resolution  $\Delta E$ of the X-IFU detector that is expected to fly on board of Athena \citep{Athena}. 
Their values are listed in table \ref{tab:instrument_specific}.
We notice that the angular size of the pixel (5''), is smaller than the
typical angular size of a few arcminutes of the typical  WHIM emitter \citep{WHIM-Takei+11}, potentially allowing one to probe the spatial and the clustering properties of the WHIM (that we model in sections \ref{subsec:results_3D_2PCF} and \ref{subsec:results_angular_2PCF})
and to efficiently remove point-like X-ray sources, like AGNs, that otherwise would artificially enhance the noise level.

The angular resolution sets the mesh size of the grid which, 
at the redshift of the output $z$, determines the number of grid points  
\begin{equation}
    N_\mathrm{grid}^3 = \texttt{int} \left(\frac{L}{\theta_\mathrm{res} \ \chi(z)}\right)^3 \, ,
    \label{eq:grid_size}
\end{equation}
where $L$ is the box size and $\chi(z)$ is the comoving distance to redshift $z$ in the assumed cosmology.

The energy resolution sets the thickness of each slice  of the data cube along the line of sight direction. For simplicity we assume the distant observer approximation, i.e. the same line of sight to all pixels in the grid. Given the small size of the simulation boxes this assumption is justified for all but the smallest redshift snapshots. For a line emission with rest-frame energy $E$ the thickness $\Delta L$ is:
\begin{equation}
    \Delta L = \frac{\Delta E}{E} \ (1+z) \ \frac{c}{H(z)} \, ,
    \label{eq:thickness_slice}
\end{equation}
where $H(z)$ is the Hubble constant at the redshift of the snapshot $z$.
This is the thickness of the surface brightness maps, i.e. 
each map is obtained by integrating along each line of sight the line flux at the positions of all gridpoints within $\Delta L $.

Since we are interested in two lines, OVII and OVIII, but would like to use the same thickness for both emission maps, we set $E=0.6$ keV, i.e. the average of the 
 OVII and OVIII emission line energies.
The area subtended by the computational cube, which represents the field of view of the map, and its thickness depends on the cosmological model adopted line each simulations and, of course, on the redshift of the snapshot. However, the dependence on the cosmology is rather mild for two reasons.
The first one is that we restrict our analysis to the low-redshift Universe. The second one is that the range of $\sigma_8$ and $\Omm$ values in the simulations used in our analysis is quite limited. Therefore, in our analysis we ignore the dependence on the cosmology and assume that the field of view and depth of all surface brightness maps are identical and equal to that of the fiducial cosmological models. Their values are listed in Table~\ref{tab:snapshot_grid}.
We further notice that detectors with an energy resolution of a few eV will produce brightness maps of substantial thickness, comparable to the size of the simulation boxes.

\subsection{Summary statistics}
\label{subsec:statistics}

Surface brightness maps generated by the integral field units contain a wealth of information that needs to be efficiently compressed.
Two summary statistics that are widely used and that we consider in this work are
\begin{itemize}
    \item the $\logNS$, i.e. the number of detected emission or the 1-point probability distribution of the pixel counts as a function of the 
line surface brightness, $S_B$.
    \item the angular 2-point correlation function, $w(\theta)$, of the line surface brightness in two pixels separated by an angle $\theta$. 
\end{itemize}

Angular cross-correlation can also be estimated if the angular position of other extragalactic sources in the same region is also available. One such example is the angular cross-correlation between the line surface brightness and the galaxies.

To characterize the WHIM properties in our simulated maps we use these statistical tools. For the latter, i.e. the angular cross-correlation, we consider the dark matter halos extracted from the simulations to avoid the complication of the galaxy-halo relation in the simulated data.

Projection effects related induced by the energy resolution of the integral field units makes it impossible to probe the spatial distribution of the line emitting gas along the sight line.
Which, however, is accessible in the simulation box. Therefore, we decided to also estimate the 2-point correlation function of the 3D surface brightness fields in the simulation as well as the cross-correlation between halo number density and the surface brightness fields. This quantity, that cannot be observed with currently proposed 
integral field units, is nevertheless useful to study the spatial distribution of the WHIM. Moreover, since in these maps the redshift of the emission line is used as a proxy to the distance of the emitter,
the estimate of this 2-point correlation function is affected by the so-called redshift distortions, i.e. from the anisotropy induced by the peculiar velocity of the emitting gas with respect to the Hubble flow
\citep{Kaiser+87,hamilton}.
Anisotropic correlation functions are characterized by nonzero quadrupole and hexadecapole moments that can be measured and used to investigate the dynamics of the line emitting gas.
For this reason, for the 3D fields, we estimate both the monopole and the quadrupole moments of the correlation function. 

To summarize, the statistical quantities that we use in this work are:
\begin{itemize}
    \item The oxygen line emitters counts $\logNS$. To estimate this quantity we simply count the number of pixels, in the line surface brightness maps, with a surface brightness level larger than $S_B$. This way we estimate the cumulative probability distribution of the OVII and OVIII line emitters. We also estimate the same quantity for both emitters, OVII and OVIII, by considering the pixels in which the OVII and OVIII surface brightness are both larger than $S_B$.
    This quantity is estimated at different redshifts using all available snapshots.
    
    Moreover, we compare our results with those of \citet{WHIM-Takei+11} and \citet{Roncarelli+12}. In their case, however, the $\logNS$ statistics is estimated considering a full lightcone to $z=0.54$ and considering angular resolutions of a few arcminutes, i.e. over larger pixels. 
    The procedure used to build a similar lightcone and to reduce the angular resolution of our maps is described in details in section \ref{subsec:lightcone}.

    \item The monopole and quadrupole moments of the spatial 2-point auto-correlation function $\xi(\tvec r)$ of the OVII and OVIII surface brightness at the points of a cubic grid. 
    Here we ignore the effect of the energy resolution of the detector and estimate the 3D correlation on the $N_\mathrm{grid}^3$ points of the grid, where the value of the $N_\mathrm{grid}$ is specified in Table 2.
    Since we are using computational boxes that assume periodic boundary conditions we use the estimator proposed by \cite{Taruya_estimator} as implemented in the \textsc{Pylians3} package.
    Input to this estimator is the fluctuation of the line surface brightness with respect to the mean,  specified on a cubic grid $\delta(\tvec x)$.
    The field is Fourier transformed, its Fourier coefficients $\delta(\tvec k)$ squared, and the result transformed back. The quadrupole moment is estimated by 
    weighting the cubic grid for $5/2 \ \mathcal L_2(\mu)$, where $\mathcal L_2$ is the 2nd-order Legendre polynomial and $\mu$ is the cosine angle between the line-of-sight and the wavelength vector $\tvec k$, before averaging over all pixel pairs.
    
    The spatial 2-point correlation functions are estimated at the redshifts of the simulations' snapshots, for the various fluctuations field defined on cubic grids. The number of gridpoints is set by the X-IFU angular resolution and, as shown in Table~\ref{tab:snapshot_grid}, is very large.
    Estimating the 2-point correlations using so many points would be computationally prohibitive and not very informative, since it would probe scales smaller than the typical halo size. Therefore, for the spatial 2-point correlation functions, we decided to re-bin the fluctuation fields on a coarser grid of $512^3$ points, corresponding to a spatial resolution of 49 kpc/$h$; a scale comparable to the size of a Milky Way sized halo.

    \item The monopole and quadrupole moments of the spatial 2-point cross-correlation function of the oxygen line surface brightness and the halo counts.
    The procedure is similar to that of the auto-correlation except for the fact that now two fields are used, one of which is obtained by interpolating the halo number counts at the positions of the $512^3$ cubic grid, and that both are fed into the Taruya estimator. 

   \item The angular auto- and cross-correlation function of the line emitters and dark matter halos, $w(\theta).$
   Also in this case the procedure to estimate this quantity is similar to the one used to compute $\xi(\tvec r)$ except for the fact that now the fluctuations fields are specified at the point of a square, rather than cubic, grid.
 \end{itemize}


\section{Results}
\label{sec:results}

In this section we present the main results of our analysis.
First of all, we visually inspect the OVII and OVIII surface brightness maps 
obtained at different redshifts and for different choices of $\Omm$ and $\sigma_8$ 
parameters and  energy feedback models.
We compute the $\logNS$ statistics statistics of these maps and evaluate their uncertainty.
Then we use the full simulation boxes to measure the spatial 2-point correlation functions of the oxygen emitters and their cross-correlation with the dark matter halos.
Finally, we estimate the angular auto- and cross- 2-point correlation functions of the above quantities from using the 2D surface brightness maps with energy and angular resolution matching these of the X-IFU detector.

\subsection{Surface brightness maps}
\label{subsec:results_maps}

\begin{figure*}
    \centering
        \includegraphics[width=.6\textwidth]{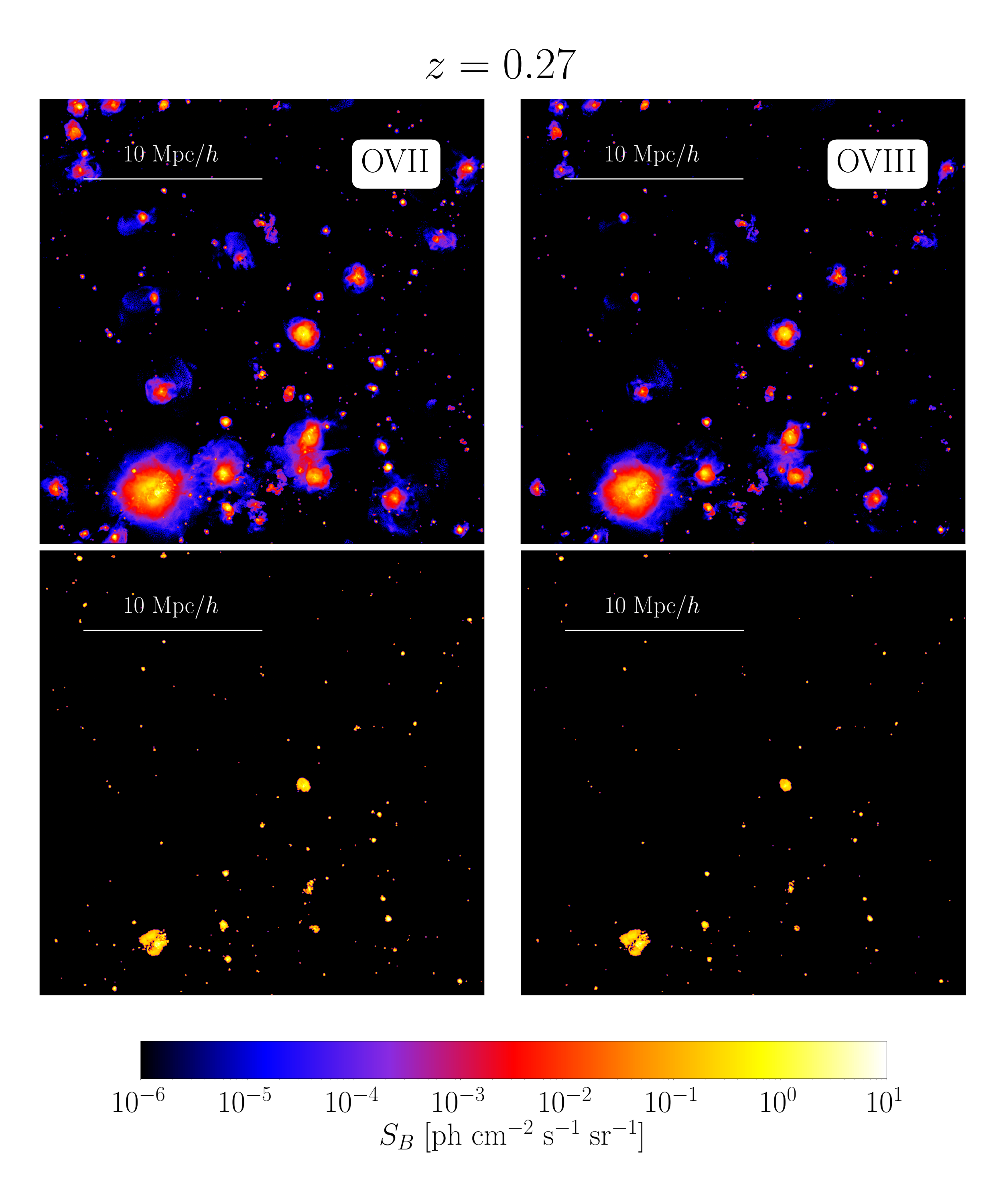}
    \caption{Surface brightness maps of the OVII (left) and OVIII (right) emission lines from the gas in slice 13.8 Mpc/$h$ thick and 25 Mpc/$h$ wide extracted from the $z=0.27$ snapshot of the \IllP simulation with the parameters of the fiducial cosmological model adopted in this work. Top panels: emission from the WHIM gas particles. Bottom panels: only pixels brighter than $0.1 \ \unitsSB$, corresponding to a 3-$\sigma$ detection threshold with a 100 ks Athena X-IFU observation are shown.}
    \label{fig:OVII-OVIII}
\end{figure*}

\begin{figure*}
    \centering
        \includegraphics[width=.8\textwidth]{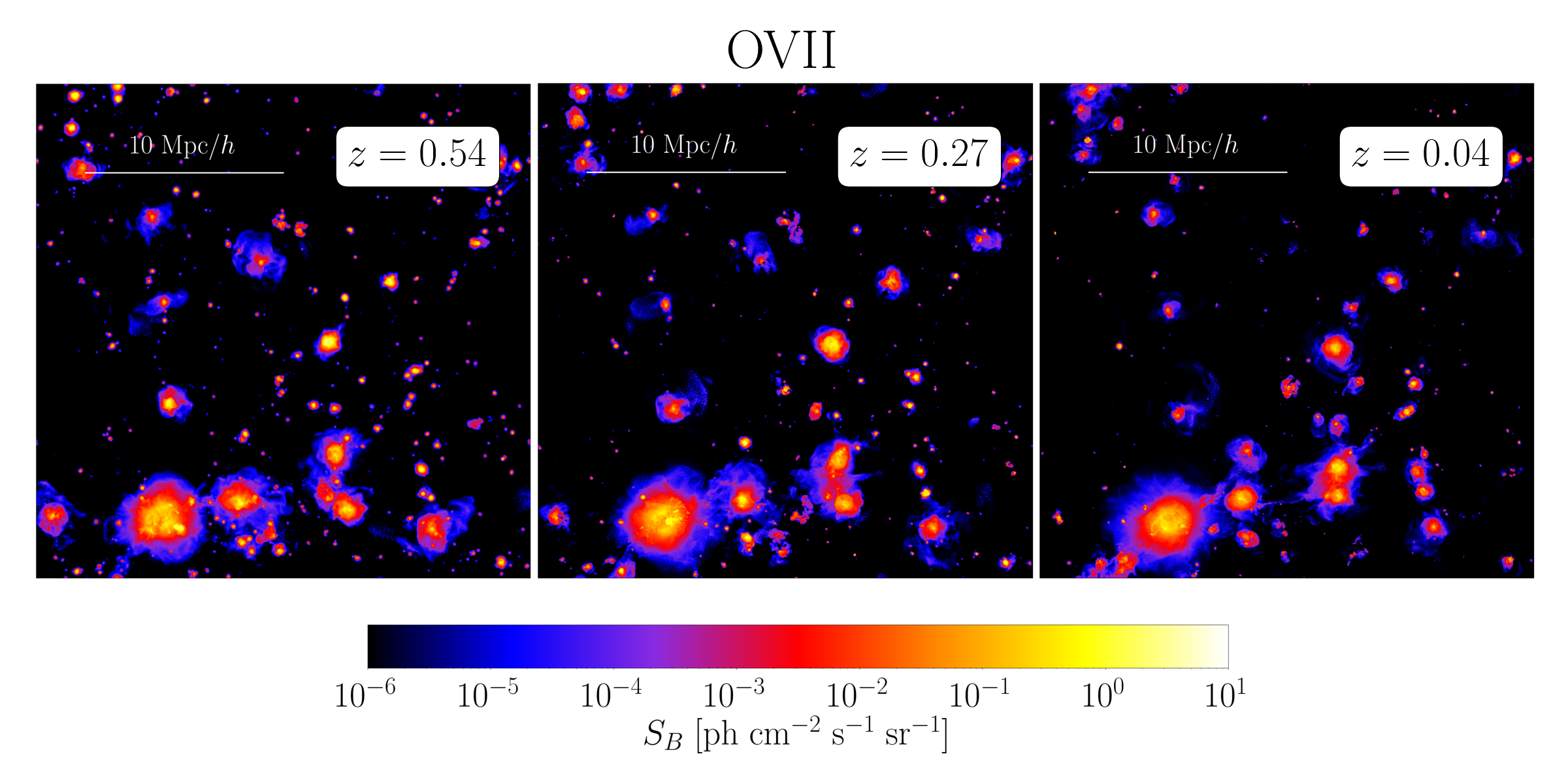}
    \caption{OVII line surface brightness maps of the same region as Fig.\ref{fig:OVII-OVIII} seen at three different redshifts: $z = 0.54,0.27,0.04$ (from left to right). The thickness of the slide changes with redshift as indicated in Table~\ref{tab:snapshot_grid}.}
    \label{fig:redshift_evolution}
\end{figure*}

\begin{figure*}
    \centering
        \includegraphics[width=.95\textwidth]{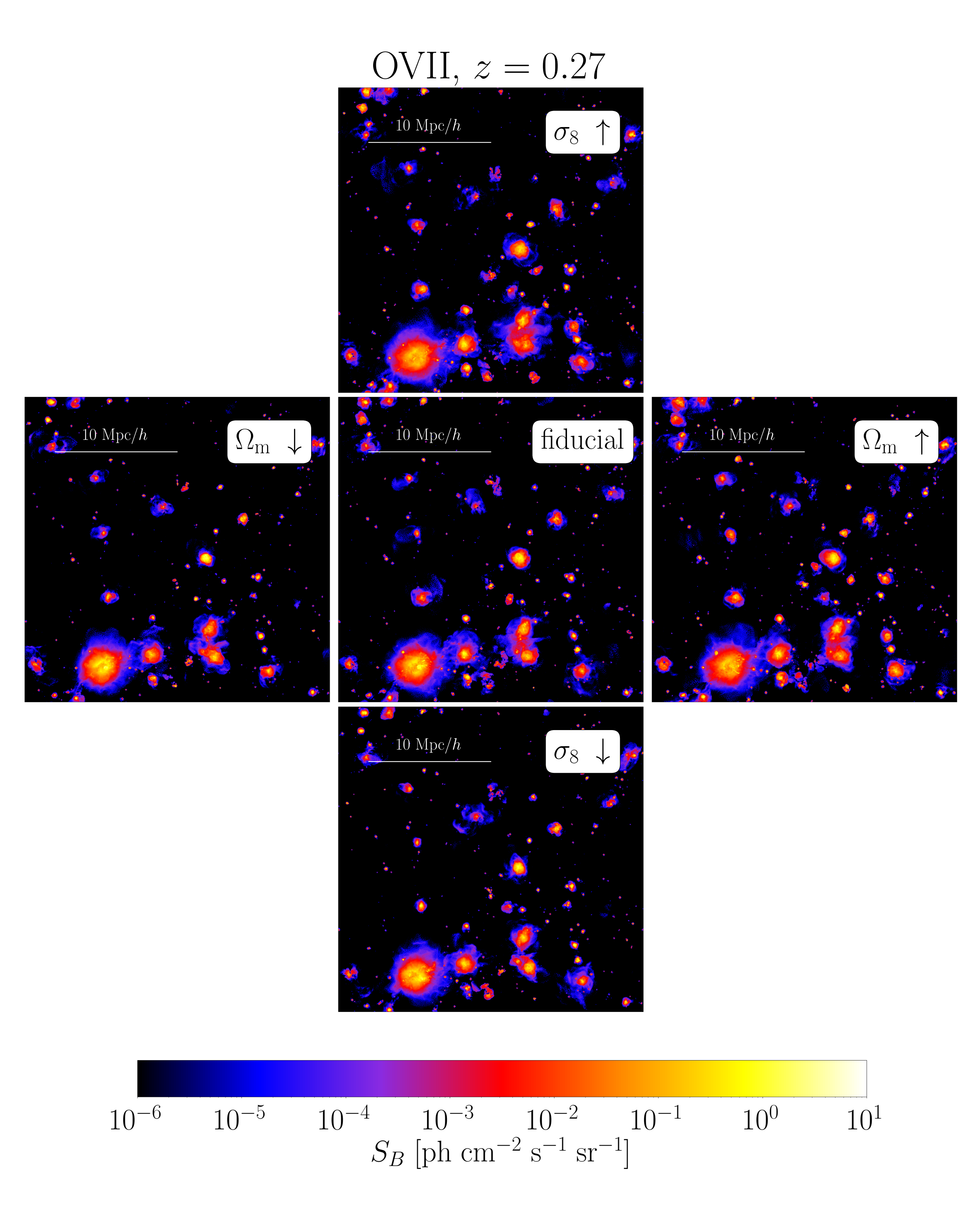}
    \caption{OVII line surface brightness maps of the same region as Fig.\ref{fig:OVII-OVIII} from the $z=0.27$ snapshot of \IllP simulations that use different values of 
     $\Omm$ and $\sigma_8$. Top to bottom: $\sigma_8$ is increased (top) or decreased (bottom) by a factor 0.04 with respect to the fiducial value $\sigma_8=0.8$ (at the center).
     Left to right: $\Omm$ is increased (top) or decreased (bottom) by a factor 0.04 with respect to the fiducial value $\Omm=0.3$ (at the center).}
    \label{fig:OVII_cosmo}
\end{figure*}

\begin{figure*}
    \centering
        \includegraphics[width=.8\textwidth]{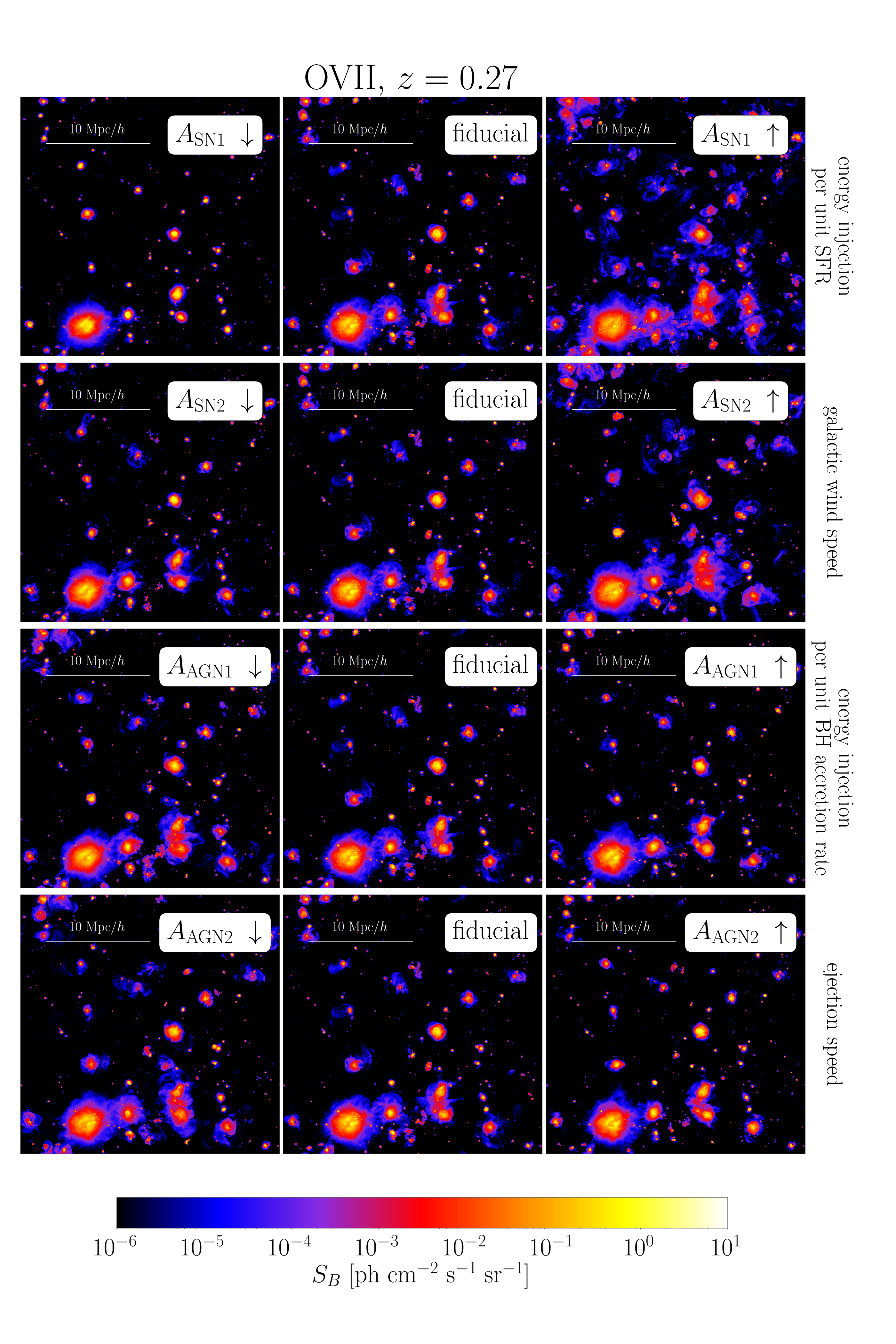}
    \caption{OVII line surface brightness maps of the same region as Fig.\ref{fig:OVII-OVIII} from the $z=0.27$ snapshot of \IllP simulations that use different values of the parameters that regulate the energy feedback. The effect of increasing each one of the four parameters, specified in the white labels, from their minimum (left panel) to the maximum (right) value, 
    with respect to the fiducial case (center) is shown in each row.
    }
    \label{fig:OVII_astro}
\end{figure*}

Fig.~\ref{fig:OVII-OVIII} shows the OVII (top left panel) and OVIII (top right) surface brightness maps of the same redshift slice taken from  the $z=0.27$ snapshot of the \IllP simulation. The parameters of the model are the same as the fiducial ones used in the \SIMCV and \IllCV sets.
We use the \IllP simulations, which have the same initial conditions, to assess the sensitivity to the choice of cosmological parameters and energy feedback models free of cosmic variance.
We first show the $z=0.27$ case since it roughly corresponds to the mean redshift in the range we have considered.
It is worth noticing that the \IllP realization is not typical. It contains one massive halo of $M\sim 9\times 10^{13} \ M_\odot/h$, the features prominently in the bottom left part of the map, and a rather large void, both quite unlikely in a random box of 
25 Mpc/$h$. 
As a result, the properties of the WHIM in this box are not representative of that in the Universe. We shall address this issue in the next sections, in which we quantify the impact of the cosmic variance on the summary statistics. Here, however, we are interested in qualitatively comparing the {\it relative} differences among different WHIM models, which requires using the same realization, however unlikely.

The surface brightness levels are color-coded according to the reference bar shown at  the bottom of the plot.
The main differences between the OVII and OVIII maps are evident: the OVIII signal is more concentrated near the center of the halos whereas the OVII is less peaked and more extended to the outskirsts. This is expected since the high gas temperature and density in the central regions of the halos are more favorable to the OVIII ions. Consequently the 1-halo term of the OVIII line surface brightness correlation is more prominent and more peaked than that of OVII, as we shall confirm in section \ref{subsec:results_angular_2PCF}).

These emission maps do not trace the spatial distribution of the oxygen ions which populate the filaments and the nodes of the cosmic web, which is clearly seen when plotting the column density rather than the emissivity of these ions, like for example in Fig. 1 of \cite{Nelson+18} or Fig. 2 of \cite{wijers19}. This of course reflects the fact that emissivity is proportional to the square of the gas density.

A more homogeneous comparison can be made with the emission maps of
\cite{Metal_line_emission-Bertone+09}, obtained from the OWLS simulations. 
They look qualitatively similar to ours
in the high density regions centered on massive halos ($M\gtrsim 10^{12} \ M_\odot/h$). However,  the diffuse emission from low density filaments is largely missing in our maps.
This difference likely result from having assumed collisional ionization equilibrium: a valid hypothesis in  high density and temperature environments that, however, underpredicts the OVII and OVIII ionization fraction in low density regions.
This would represent a severe limitation if our target were the study of the WHIM property in absorption. However, we are interested to study the WHIM in emission, a possibility that, with next generation instruments like X-IFU, will be limited to high density regions, where collisional equilibrium holds true.
This point is clearly demonstrated in the two bottom panels in which we show the same emission maps 
after removing all pixels with surface brightness below $0.1 \ \mathrm{photons \ cm^{-2} \ s^{-1} \ sr^{-1}}$, which represents a 3-$\sigma$ detection threshold for a 100 ks observations with Athena X-IFU \citep{WHIM-Takei+11,kaastra13}.
Clearly, the only OVII and OVIII line emission that will be detected is the one generated by gas associated to dark matter halos. Weaker emission from halo outskirts could only be detected by stacking or by cross-correlating line surface brightness maps with galaxy positions  in spectroscopic redshift surveys, if these are available in the observed areas.

The evolution of the line surface brightness as a function of time can be appreciated in Fig.  \ref{fig:redshift_evolution}, where we show the same OVII map at $z=0.27$ as in
Fig.~\ref{fig:OVII-OVIII} next to the maps at obtained from the $z=0.54$ (left) and $z=0.04$ (right) snapshots.
Line emission is always associated to halos and become progressively more concentrated around the massive ones as result of the merging process.
We only show the OVII map since the redshift dependence of the OVIII map are entirely similar.
The signature of this redshift evolution will be seen, more quantitatively, in the spatial correlation properties of the emitting gas, as we shall see in sections \ref{subsec:results_3D_2PCF} and \ref{subsec:results_angular_2PCF}.

Fig.~\ref{fig:OVII_cosmo} illustrates the sensitivity of the surface brightness map to $\sigma_8$ and $\Omm$, i.e. the two free cosmological parameters of our analysis.
The central panel shows the same map at $z=0.27$ as in Fig.~\ref{fig:OVII-OVIII}.
The effect of increasing $\sigma_8$ can be appreciated from bottom to top, whereas that of increasing $\Omm$ from left to right. Both parameters are increased/decreased by the same amount, 0.04, with respect to the fiducial case.
In creating these maps, we have kept the comoving depth of the slice fixed to the one of the fiducial cosmology ignoring the small dependence on $\Omm$. 
The effect of increasing the value of $\Omm$ is that of shifting the emission from the centers to the outskirts of the halos, especially those with $M\lesssim 10^{12} \ M_\odot/h$.
A similar effect, though less prominent, is obtained when increasing   $\sigma_8$.  
This result is not surprising: increasing the value if either parameters increase the clustering amplitude and the merging rate.
As a result the central region of the halos becomes so hot and dense that the ion fraction of OVII and even OVIII drops.
while the merging process accretes warm gas located in the outer part of the halos.
The similarity of the response to the change in
$\sigma_8$ and  $\Omm$ implies a certain degree of degeneracy in determining the physical properties of the line emitting gas.
Also in this case we omit to show the OVIII maps since the results are  very similar.
 
\begin{figure*}
	\includegraphics[width=.8\textwidth]{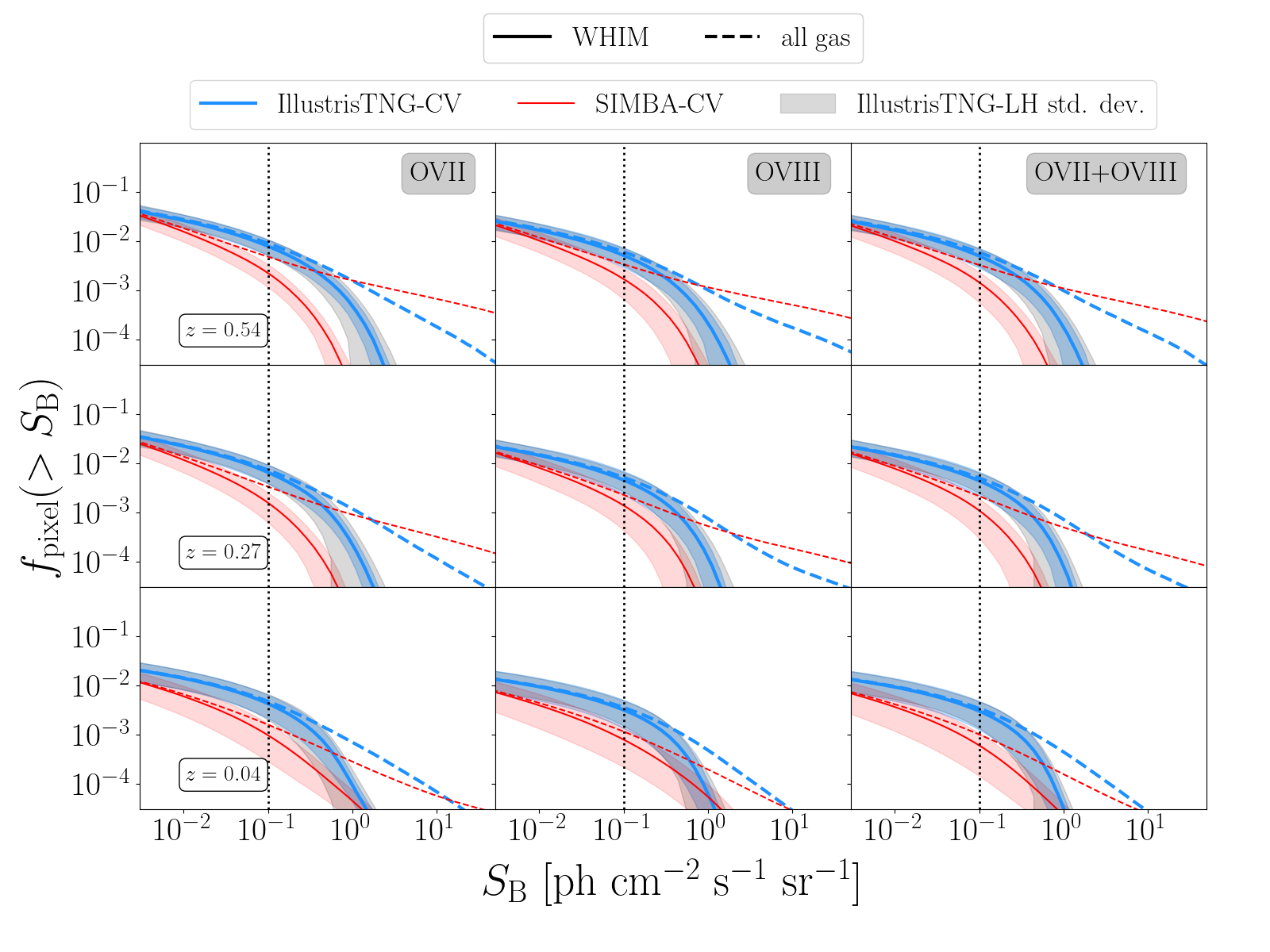}
    \caption{
    $\logNS$ for OVII (left), OVIII (center) and OVII+OVIII (right) line counts in the brightness maps at three different redshifts: 0.54 (top), 0.27 (center) and 0.04 (bottom).
    In this and following figures we will always plot the \textit{fraction}, rather than the total number, of pixels above a given $\SB$ threshold, $f_\mathrm{pixel}(>S_B)$.
    Cyan, thick, continuous curves and shaded area show the WHIM line counts    averaged among the 27 \IllCV realizations and their {\it rms} scatter.
    Red, thin, continuous curves and shaded area show the same quantities for the \SIMCV case.
    Dashed curves of the same colors show the average line counts generated by all gas particles, including the WHIM.
    The grey band surrounding the  \IllCV WHIM counts shows the {\it rms} 
    scatter among the 63 \IllLH realizations.
    All the $\SB$ left to the vertical dotted line at $0.1 \ \unitsSB$ are below the 3-$\sigma$ detection threshold in a 100 ks observation with Athena X-IFU.
    }
    \label{fig:CDF_CV_set}
\end{figure*}
 
The effect of varying the four parameters that characterize the energy feedback models, 
illustrated in Fig.~\ref{fig:OVII_astro}, is more spectacular. The four rows show the effect of increasing the parameters one by one, from left to right, with respect to the fiducial case, shown in the central panels. The largest effect is obtained when changing the amount of energy injected through SN explosion ($\ASN{1}$ parameter, top row) and the speed of the galactic winds ($\ASN{2}$ parameter, second from the top). Increasing both effects significantly enhances the surface brightness in halo surroundings without dimming that in the center.
The fact that increasing the amount of energy injected increases the line emissivity of the intergalactic medium is not surprising and confirms the results of  \cite{Roncarelli+12} whom, however, used a different hydrodynamic simulation and energy feedback criteria.

The effect of energy feedback from AGNs, regulated by the two parameters $\AAGN{1}$ and $\AAGN{2}$, shown in the two bottom rows, is comparatively less prominent.
This behavior is not unexpected. SNe go off in halos of all masses while AGN feedback is more efficient in massive halos, which  are rare in the limited volume of the \IllCV
simulations. As a result, the AGN feedback mechanism is typical subdominant in the CV set.

Due to the limited volume of the CV set, halos more massive than $ 10^{13} \ M_\odot/h$ are rare, 
so that AGN feedback is important only for a minimal fraction of them.
It also goes in the opposite direction: increasing the energy release from AGN increases the surface brightness near the center of the halos but decreases it in the outer parts.
Finally, we notice that there must be a strong interplay among the different astrophysical parameters: for instance, increasing SN feedback will make it harder to create massive galaxies, i.e. the ones that are most likely to host an AGN; similarly, increasing AGN activity may release more energy and make SN explosions more or less efficient.
The competing effect of these two energy release mechanisms on the OVII (but also OVIII) surface brightness is an interesting effect that we will quantify in the next section using summary statistics.

\subsection{$\logNS$ of the OVII and OVIII line emitters}

\label{subsec:results_CDF}

The simplest, commonly used, 1-point statistics that we consider is the $\logNS$ of the oxygen line emitters that we estimate by 
counting the number of pixels $N$ in the maps above with a line surface brightness above $S_B$.
The results are shown in Fig. \ref{fig:CDF_CV_set} for the OVII (left), OVIII (center) and for the joint OVII+OVIII (right) counts at three different redshifts ($z=0.54,0.27,0.04$, from top to bottom).
Solid curves show the fraction of pixels in the maps in which the surface brightness of the lines emitted by the WHIM gas only is above $S_B$, indicated in the X-axis. 
These curves represent the average counts over the available 27 maps obtained from the  \IllCV  (cyan) and 
\SIMCV (red) simulations.
The surrounding shaded areas of the same color shows the {\it rms} scatter among the 27 maps and represent the standard deviation of the counts. Since the parent simulations differ only by the choice of the initial conditions they quantify the impact of the cosmic variance.
The broader grey band, plotted for the  \IllCV case only to avoid overcrowding shows the \textit{rms} scatter estimated from the 63 \IllLH realizations characterized by 
different  initial conditions and model parameter values. The latter varies within the ranges specified in Table~\ref{tab:simulation_sets_table}. 
Therefore the grey band quantifies the impact of cosmic variance and model parameter uncertainties. At small $S_B$ values the error budget is dominated by cosmic variance since the blue and grey bands coincide. At large $S_B$ values errors are dominated by model uncertainties, especially by those associated to the energy feedback models that are  weakly constrained by observations.
Dashed curves show the $\logNS$ curves of the oxygen lines generated by all the gas particles, not just the WHIM.
Finally, a vertical line is drawn in correspondence to the expected minimum surface brightness required to detect oxygen lines at 3-$\sigma$ significance with a 100 ks observations with Athena X-IFU.

Perhaps the most interesting result is the systematic difference between the $\logNS$ predicted by the SIMBA simulations and the IllustrisTNG ones.
A difference in the simulation outputs is to be expected. What is remarkable is the statistical significance of this difference which is of the same order as the statistical uncertainties quantified by the colored bands. Therefore, differences in the techniques used to perform hydrodynamic are the source of systematic errors that cannot be ignored in the error budget.
More specifically the SIMBA simulations systematically underpredict the number of WHIM emitters with respect to IllustrisTNG.
This is also the case for the weak emission from the whole gas, but not for the strongest emitters, whose number is larger in the SIMBA case at low redshift.
Another interesting feature is the high similarity between the OVIII and OVIII+OVII curves, at all redshifts, for all type of gas particles and for both hydrodynamic experiments. In practice, the detection of an OVIII line almost guarantees that an OVII line is also observed, presumably from the same emitter, hence providing an unambiguous determination of the redshift of the emitter. For this reason, most often than not, we will only show the OVIII $\logNS$ curves in the rest of the article. 
This result confirm that of \cite{WHIM-Takei+11} based on yet another set of hydrodynamic simulations, though in their case the similarity between the OVIII and OVII+OVIII $\logNS$ counts was less striking than here.

Finally, given the large number of pixels in the maps, $N_\mathrm{grid}^2$, one expects a large number of 3-$\sigma$ detections with Athena X-IFU, many of them associated to the same physical structure. We shall return on this point and estimate the number of expected emitters per line-of-sight in section \ref{subsec:lightcone}.

The behaviour of the continuous  curves of Fig. \ref{fig:CDF_CV_set} shows a characteristics break or knee that is qualitatively similar to that of the column density distribution function of the OVII and OVIII absorbers in the EAGLE and Illustris simulations \citep{wijers19}.
This is not surprising since X-ray oxygen absorbers trace the bulk of the WHIM but avoid the hotter and denser regions where the strongest emission lines are generated. These lines are included in the counts that consider all gas particles, and not just the WHIM, represented by the 
dashed curves of of Fig. \ref{fig:CDF_CV_set} that show no clear break.

\begin{figure*}
	\includegraphics[width=.9\textwidth]{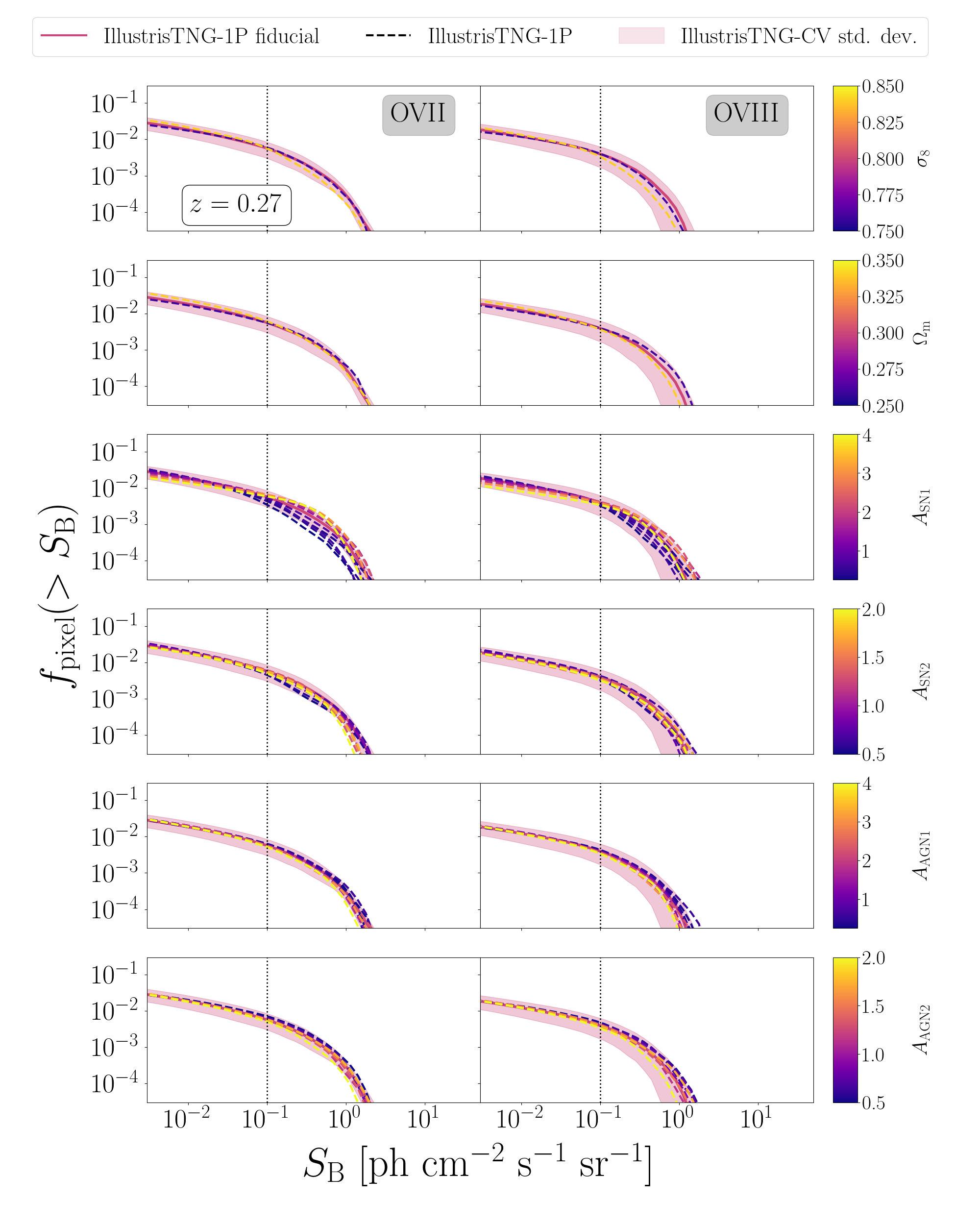}
    \caption{$\logNS$ statistics for OVII (left panels) and OVIII (right panels) line counts in the \IllP  subset. In each row the variation of a single parameter with respect to the fiducial model is shown, color-coded according to the parameter value. The \IllP prediction is surrounded by a purple shaded area representing the standard deviation as measured from the {\it rms} scatter among the 27 realizations of the \IllCV  subset.
    The vertical black dotted line identifies the expected 3-$\sigma$ detection threshold of a 100 ks observation with Athena.}
    \label{fig:CDF_CV_LH_1P_sets}
\end{figure*}

\begin{figure}
    \includegraphics[width=\columnwidth]{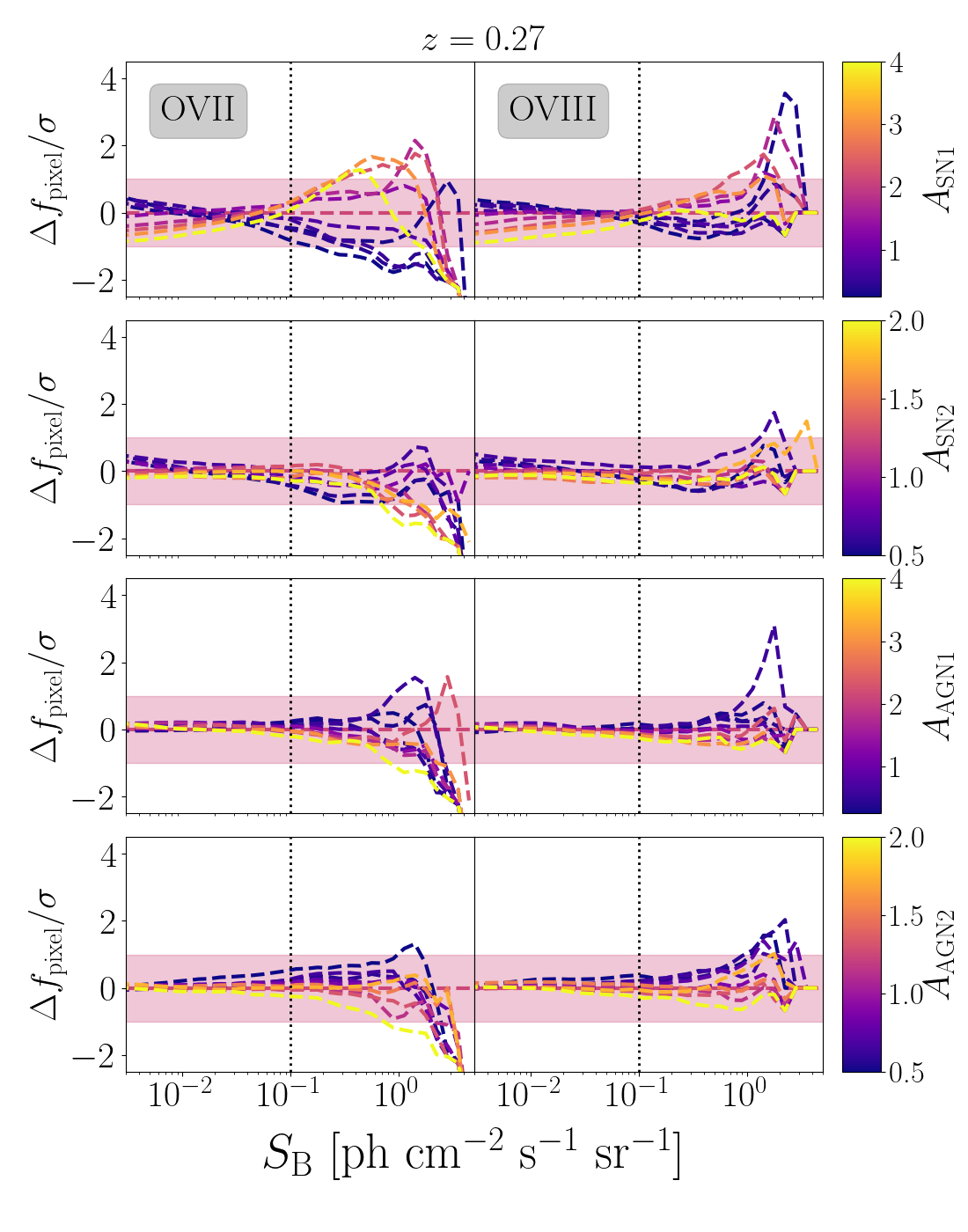}
    \caption{This figure reproduces the lower four panels of Fig. \ref{fig:CDF_CV_LH_1P_sets}, where we show $\logNS$ when varying the feedback parameters.
    Here however we highlight the relative differences of $\logNS$ with respect to the fiducial model in units of the statistical uncertainty measured from the CV set.}
    \label{fig:CDF_CV_LH_1P_sets_ratio}
\end{figure}

Fig.~\ref{fig:CDF_CV_LH_1P_sets} illustrates the sensitivity of the OVII and OVIII WHIM line counts to  the six parameters of the model with respect to the fiducial case (represented, in each panel, by the continuous purple curve). The surrounding shaded area shows the \textit{rms} scatter among the 27 \IllCV realizations.
We only show the $z=0.27$ snapshots since the qualitative dependence on the different parameters is the same at all redshifts.
Dashed curves show the $\logNS$ obtained when changing one parameter at the time, one in each row. Parameter's values are color-coded and indicated in the vertical color bar.

The first two rows show that the model $\logNS$ are quite insensitive to the choice of $\sigma_8$ and $\Omm$, when these are changed in the allowed intervals that generously brackets the one constrained by observations. Some difference is seen in the counts of the very bright emitters. It quantifies the results of the qualitative inspection of the brightness maps in Fig.~\ref{fig:OVII_cosmo} due to the competing effect of these two parameters on the gas mass accretion and halo merging histories. Variations, however, are well within the expected cosmic variance in the observed region.

The effect of changing the energy feedback parameters is much more evident, as we have qualitatively seen in the surface brightness maps shown in Fig.~\ref{fig:OVII_astro} 
where differences in the bright line counts with respect to the fiducial case can be larger than the cosmic variance.
To better appreciate their effect we plot, in Fig.~\ref{fig:CDF_CV_LH_1P_sets_ratio}
the residual line counts with respect to the fiducial case in units of cosmic variance.

Among all parameters the one that has the largest impact is the one that controls the energy injection into the intergalactic medium due to star formation processes.
Increasing its value systematically increases the number emission lines with surface brightness as high as $\sim 1 \ \unitsSB$ (and vice versa).
The effect is (barely) larger than the expected cosmic variance for the 25 Mpc/$h$ box.
For brighter emitters the trend seems to be the opposite. The actual significance of this inversion, however, is questionable since in this very low number count regime the Gaussian hypothesis does not hold anymore, and the measured \textit{rms} scatter is not sufficient to characterize the statistical significance of the event.
The effect of the other three parameters is smaller and the trend less clear. Increasing the energy feedback from AGNs, regulated by the two parameters $\AAGN{1}$ and $\AAGN{2}$, decreases the number of bright counts, hence confirming the qualitative analysis of the OVII surface brightness maps, though the variations are comparable to the cosmic variance.

The limited number of WHIM counts that drops to zero above $\sim 3 \ \unitsSB$ 
makes it difficult to analyze the very bright regime $\geq 1 \ \unitsSB$, where the effects of the energy release processes are expected to be larger. However, one can look further into this regime by relaxing the constraints that arbitrarily define the WHIM gas and consider emission from particles with \textit{total} (rather than gas) mass overdensity smaller than 1,000. 
Doing so allows us to probe regions of higher density, in which AGN and stellar feedback processes are strong enough to change the density of the gas but not that of the dark matter.
The resulting $\logNS$ counts and their trend are qualitatively similar to those of \ref{fig:CDF_CV_LH_1P_sets_ratio}. However, the difference with respect to the fiducial model in the bright counts regime are much larger, boosting up to $6-8$ times the cosmic variance level.

\subsection{The spatial auto and cross 2-point correlation functions}
\label{subsec:results_3D_2PCF}

\begin{figure*}
	\includegraphics[width=.8\textwidth]{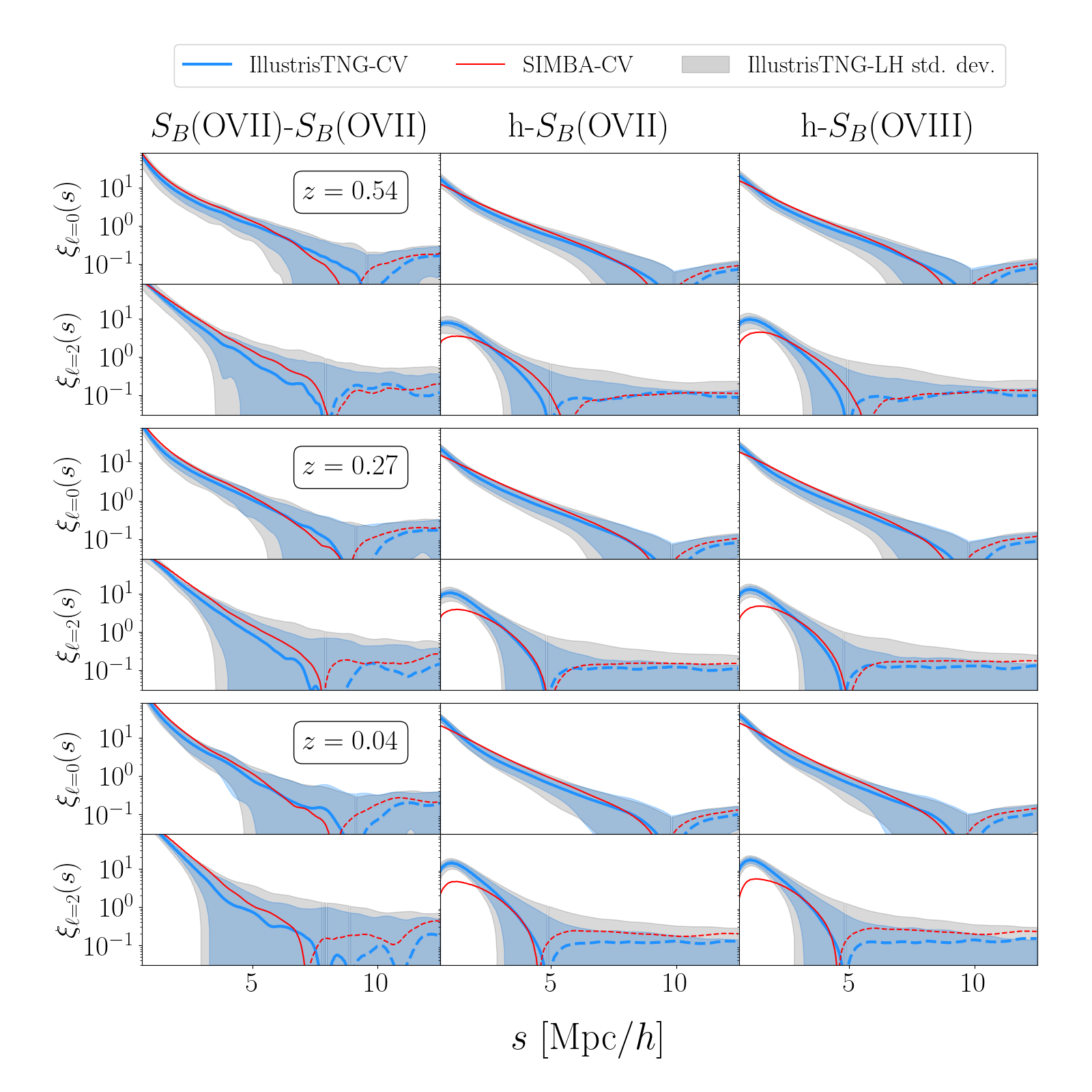}
    \caption{Three-dimensional auto- and cross- 2-point correlation function of WHIM line emission and halos. The figure is divided in three part. The top and the bottom plots of each one show the 
    monopole and quadrupole moments of the OVII line $\SB$ auto-correlation (left panels) and halo-$\SB$ of OVII and OVIII cross-correlations (central and right panels, respectively) at $z=0.54,0.27,0.04$ (top, middle and bottom row, respectively). Cyan (red) curves show the average measurement over the
    27 \IllCV (SIMBA-CV) realizations. Solid lines indicate positive correlation. Dashed line is for negative correlation.  Cyan and grey shaded areas quantify the {\it rms} scatter among the 
    27 and 63 \IllCV and \IllLH realizations, respectively.}
    \label{fig:3D_2PCFs_CV}
\end{figure*}

In the surface brightness maps the intensity of the oxygen line emission peaks in correspondence to the position of  the dark matter halos. This association reflects an underlying spatial correlation that one cannot measure directly because of the finite energy resolution of the current
X-ray imaging spectrometers. However, nothing prevents us from doing so in the simulations.
We show in Fig. \ref{fig:3D_2PCFs_CV} the monopole and quadrupole moments of
the 2-point auto-correlation function of the OVII line surface brightness fluctuations (left column) and the cross-correlations function of OVII (center) and OVIII (right) line surface brightness fluctuations and the 
fluctuations in the number density of the halos.
All these quantities are defined at the $N_\mathrm{grid}=512^3$ gridpoints in the simulation box and only emission from the WHIM gas has been considered.
Results are divided in three rows, each one showing a different redshift snapshot indicated in the labels. For each redshift we show both the monopole (upper panel) and quadrupole (lower panel) moments.

In the plots we use the same color code as in the previous figures. 
The \IllCV results are similar to the \SIMCV ones at all redshifts, except for the quadrupole moment of the cross-correlation at very small separations. This is remarkable, considering the systematic differences in the $\logNS$ predictions. In other words the two type of simulations predict different number of emitters but with similar spatial correlation properties. 
In all cases explored the monopoles of the auto- and cross-correlation functions have the familiar power-law shape for separations below $\sim 7$  Mpc/$h$), followed by a transition to negative values.
This transition is an artifact induced by the so called integral constraint, i.e. the fact that 2-point function integrated over the sample volume must vanish. The small size of the simulation box places this transitions at separations much smaller than in the typical clustering analyses over cosmological volumes. As the correlation amplitude at small separation increases with the redshift, the transition radius correspondingly shifts towards smaller separations.

The integral constraints does not apply to the quadrupole moment. That is to say that the zero crossing, in this case, is a genuine feature that marks the transition between small scales characterized by incoherent motions in high density structures and large scale coherent matter infall towards high density regions.
The zero crossing scale decreases with the redshift, which indicates that incoherent motions becomes progressively restricted to more compact virialized structure as a result of the halo merging process, while the WHIM gas is coherently accreted onto the nodes and the filaments of the large scale structure.
Interestingly, the zero crossing for the cross-correlation occurs at smaller separation than for the OVII auto-correlation.  This suggests that the relative motion of the collisional gas particles are typically less coherent than that of the gas with respect to the collisionless halos.

Finally, we note the similarity between the OVII-halo and OVIII-halo cross-correlation functions (and consequently between the OVII-OVII and OVIII-OVIII auto-correlation, not shown). 
This is a bit surprising, since OVIII emitters are expected to preferentially populate virialized regions where the fraction of OVII ions is lower.
Yet this result is robust, since it is obtained from both simulation sets, and statically significant since, as is seen in \ref{fig:3D_2PCFs_CV}, the signal to noise is clearly boosted by the cross-correlation, in comparison to the auto-correlation case.

\subsection{The angular auto and cross 2-point correlation functions}
\label{subsec:results_angular_2PCF}

\begin{figure*}
	\includegraphics[width=.8\textwidth]{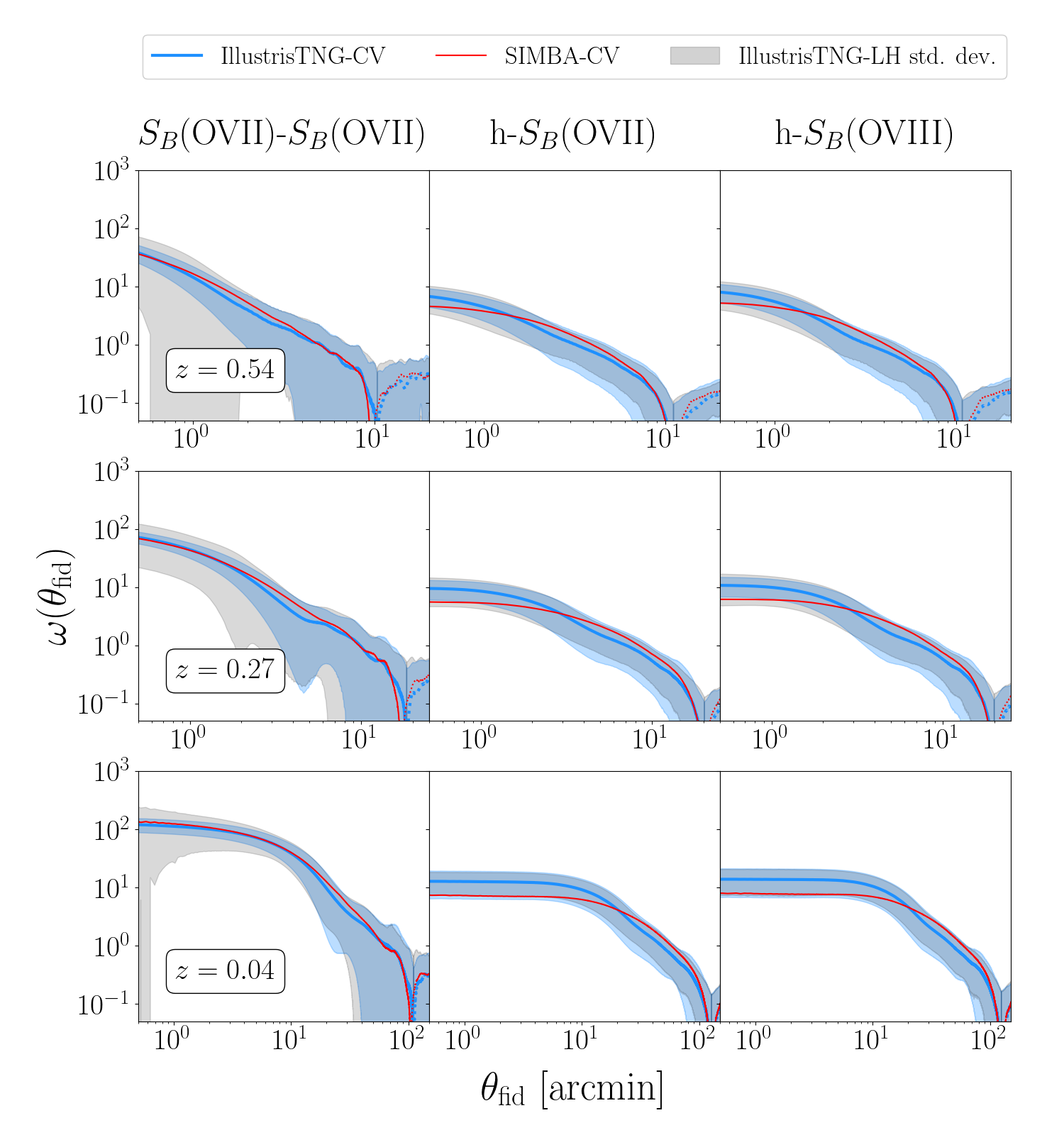}
    \caption{Angular 2-point correlation function of the OVII WHIM line surface brighntess 
    (panels on the left) and cross-correlation of the OVII and OVIII WHIM lines and  the halos in the simulations (panels at the center and on the right, respectively).
    The plots ion the three rows show the measured correlation at the redshifts of three different snapshots:  $z=0.54,0.27,0.04$ (top, middle and bottom row, respectively).
    In each panel the cyan (red) curves represent the average correlation function measured among the 27 \IllCV (\SIMCV) simulations. Cyan and grey bands show the {\it rms} scatter among the  27  \IllCV and 63 \IllLH realizations, respectively.}
    \label{fig:angular_2PCFs_CV}
\end{figure*}

Integral field spectroscopy will not allow us to trace the spatial distribution of the line emitting gas. Instead, it allows us to map the line emission within redshift slices whose thickness is determined by the energy resolution of the instrument. As a result, one will be able to perform a tomographic analysis by measuring the angular correlation properties of the surface brightness maps (and their cross-correlation) in the different slices. 

In Fig.~\ref{fig:angular_2PCFs_CV} we show the expected result of such analysis.  The 2-point auto and cross angular correlation functions shown here are the projected versions of the spatial ones shown in Fig.~\ref{fig:3D_2PCFs_CV}. We observe the same features that we have already described in the previous section: the presence of a zero crossing induced by the integral constraint, the consistency between the \IllCV and the \SIMCV predictions, and the similarity between the OVII and OVIII-halo cross-correlation functions, i.e. the fact that the projected density profile of the WHIM around the halo centers can be traced using either the OVII or OVIII lines.
Projection effects magnifies a feature that was difficult to catch in the spatial correlation function: the core-shaped profile in the inner region. 
This is very evident for the cross-correlation but also present in the auto-correlation, especially at low redshifts. A similar feature was seen in the OVII-OVIII and OVIII-OVIII auto-correlation measured in the IllustrisTNG simulation by \cite{Nelson+18} (see their Fig. 7). The likely explanation is that this core-shaped profile indicates that the fraction of the OVII and OVIII decreases when moving towards the center of dark matter halos where the temperature and the gas density steadily increase.

Although we are not attempting to provide realistic forecast to future observations, it is important to note that the signal-to-noise of the angular cross-correlation is higher than that of the auto-correlation. This result indicates that an effective strategy to statistically detect the WHIM and to study its spatial distribution in relation to that of halos is to cross-correlate the observed surface brightness maps with the angular distribution of galaxies in the same redshift slice taken from existing (or follow-up)  spectroscopic, or photometric, galaxy redshift catalogs.

\begin{figure*}
	\includegraphics[width=\textwidth]{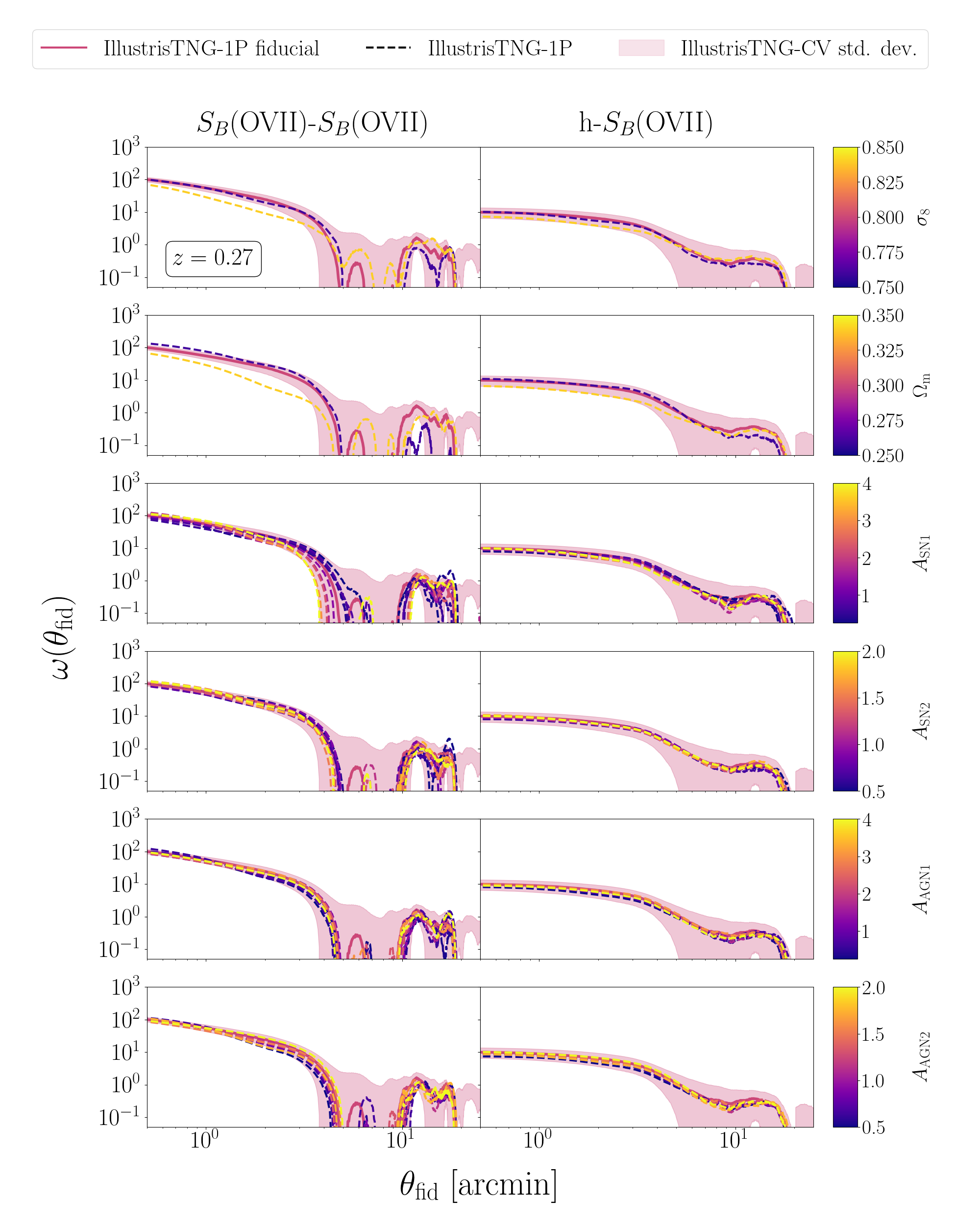}
    \caption{Angular auto- 2-point correlation function of the OVII WHIM line surface brightness (left panels) and OVII-halo cross-correlation (right panels) measured in the \IllP simulations. 
    In each row the effect of varying one parameter of the model, indicated on the right Y-axis, is shown. The color code and the meaning of the colored band are the same as as in  Fig.\ref{fig:CDF_CV_LH_1P_sets}.}
    \label{fig:angular_2PCFs_CV_LH_1P_sets}
\end{figure*}

In the last figure of this section we show how the angular correlation functions depend on the parameters of the models. We have repeated the same analysis as for the $\logNS$ and considered the 
effect of changing one parameter at the time in the \IllP simulation. 
The results are summarized in Fig.~\ref{fig:angular_2PCFs_CV_LH_1P_sets}, in which we show the angular auto-correlation of the OVII surface brightness (left column) and the halo-OVII cross-correlation (right column) for various choices of the cosmological parameters  $\sigma_8$ and $\Omm$ (upper two rows) and of the parameters that regulate the energy feedback. Color codes and line styles are the same as in Fig.~\ref{fig:CDF_CV_LH_1P_sets}.
We do not show the OVIII results since they are very similar to those of the OVII emitters.

Contrary to the  $\logNS$ case, the angular correlation properties are more sensitive to changes in the cosmological model than in the energy release mechanisms. Increasing either $\sigma_8$ or $\Omm$ has the same effect of reducing the correlation amplitude at small angular scales, in  both the auto- and the cross-correlation functions.
This seems counter intuitive, since both parameters regulate the clustering of the matter. However, a larger clustering amplitude means that structures, and halos, virialize earlier, increase their mass and temperature and reduce the OVII and OVIII ionization fractions (in favor of fully-ionized oxygen).
On the contrary, the clustering properties of the emitters are remarkably insensitive to the 
details of the energy feedback processes.
We only detect a positive correlation amplitude at small angular scales on the energy release from star formation ($\ASN{1}$) but the dependence is weak and well within the standard deviation strip.

The different sensitivity of the two statistics potentially measurable with next generation detectors, $\logNS$ and $w(\theta)$, to the cosmological and astrophysical parameters that characterize the WHIM model is remarkable.
It suggests that the combination of the two observations, together with their redshift dependence, has the potential of removing parameters' degeneracy and effectively constrain the most uncertain aspect of the WHIM models: how and where the energy release from star formation processes and AGN activity affect the evolution of the missing baryons in the Universe.

\subsection{Sensitivity to the survey volume}
\label{subsec:volume_resolution_effects}

The limited volume of the simulation box in the CAMELS suite potentially affects the properties of the WHIM.
First of all, neither the individual simulations nor, possibly, their average over the set of the 27
\IllCV and \SIMCV realizations are representative of the cosmic mean.
Second, these simulations  neglect the contribution of the large scale modes on the evolution of the large scale structures, which potentially bias the predictions of the WHIM properties. 
Finally, and consequently, summary statistics are also affected by the volume constraint. One example is the integral constraint for 2-point clustering statistics.

To assess the volume effect we have repeated the same analysis 
using the \IllBig \ simulation \citep{IllustrisTNG_simulations}.
The volume of this simulation is $\sim 550$ larger than that of the CAMELS and the mass resolution is also higher, which allows us to assess both the volume and mass resolution effects in one go.

\subsubsection{$\logNS$}
\label{subsec:volume_resolution_effectscounts}

\begin{figure*}
	\includegraphics[width=.9\textwidth]{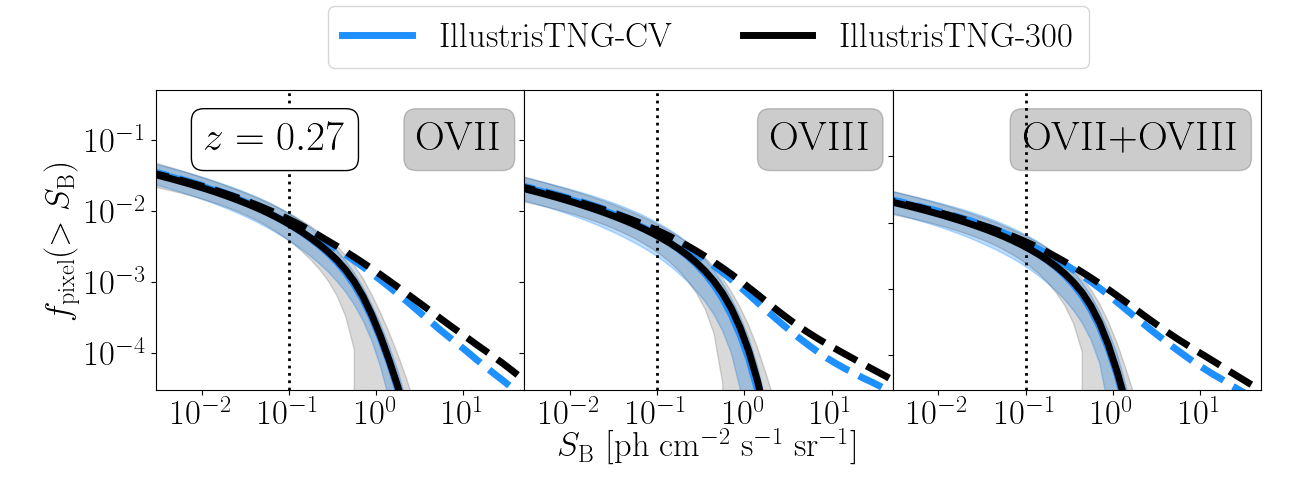}
    \caption{Comparison between the fraction of WHIM emitters in OVII (left), OVIII (middle) and both (right) for a snapshot at $z=0.27$ of the \IllBig \ (black solid lines) and the mean of the 27 realizations of the \IllCV set (blue solid lines).
    Dashed lines refer to all the gas particles.
    Blue and grey shaded areas surrounding the lines represent the scatter of the \IllCV and \IllLH sets, respectively.
    The vertical dotted black line marks the Athena sensitivity.}
    \label{fig:CDF_I300_CV}
\end{figure*}

In Fig.~\ref{fig:CDF_I300_CV}  we compare the average $\logNS$ of the oxygen line counts at $z=0.27$ in the 27 \IllCV realizations (thick solid cyan curve, form  Fig.~\ref{fig:CDF_CV_set}) with the same counts  performed in the larger \IllBig \ simulations. Continuous curves accounts for line emission for the WHIM only, while dashed curves accounts for contribution from all gas particles.
The standard deviation areas and their colors are the same as in Fig.~\ref{fig:CDF_CV_set}.
The match between the two sets of almost superimposed continuous curves is remarkable. It shows that the 27 realizations considered so far are enough to probe the statistics of the WHIM emitters and that they are not appreciably affected by systematic errors driven by volume and mass resolution effects.
Some difference can  be appreciated in the rare counts of bright lines emitted by all gas particles. In this case the average line counts in the small boxes are systematically less than in the larger \IllBig \ box, although the size of the effect is smaller than the expected uncertainty (not shown to avoid overcrowding the plots). 

\subsubsection{$w(\theta)$}
\label{subsec:volume_resolution_effectswtheta}

\begin{figure*}
	\includegraphics[width=.9\textwidth]{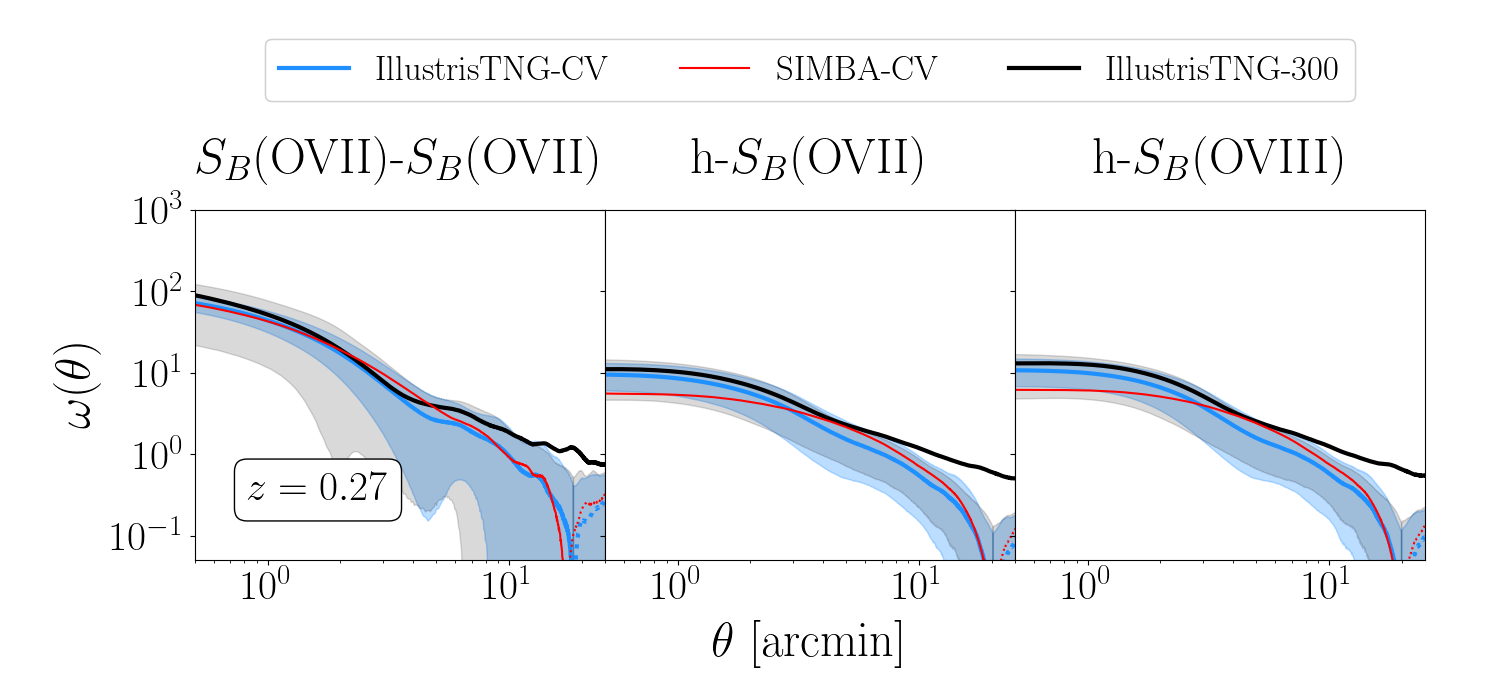}
    \caption{Comparison of the angular 2PCFs for $\SB$ of OVII (left panels) and halo-$\SB$ of OVII and OVIII (central and right panels, respectively) for the IllustrisTNG-300, \IllCV and \SIMCV simulations. We show here the snapshot at $z=0.27$. Blue lines and shaded area represent the mean and scatter of the 27 realizations of the \IllCV subset; red lines do likewise for the \SIMCV subset. The grey shaded area is the scatter of the \IllLH simulations we consider. The solid black line is the measurement from the \IllBig \ simulation.}
    \label{fig:angular_2PCFs_IllustrisTNG300}
\end{figure*}

The results shown in Fig.~\ref{fig:angular_2PCFs_IllustrisTNG300} quantify the sensitivity of the angular auto- and cross-correlation functions to the volume of the simulated box, at the same redshift ($z=0.27$) as the $\logNS$ curves (and, as in that case, we verified that the results do not depend ion the redshift considered).
The agreement between the  \IllCV and the \IllBig \ angular correlation functions 
(blue and black curve, respectively  within the uncertainty strip is evident at
small angular separation up to $\sim 10$ arcmin. In this range the shapes of the auto and cross-correlation function are almost identical, whereas the amplitude is larger for the \IllBig \ case, reflecting the fact that these simulations contain a population of larger, highly biased halos that is largely missing in the \IllCV boxes.
However, the difference in amplitude is smaller than that between the \IllCV  and \SIMCV (red curve) cases.

On larger angular scales the  \IllBig \ angular correlation steadily decreases but remains positive whereas the other curved drop to zero.
As we have already pointed out, this zero crossing is simply a manifestation of the integral constraint that, in the case of the larger \IllBig \ simulations, becomes effective on much larger scales.
As a result the zero crossing scale shifts on correspondingly larger angular separations.

\subsection{The halo-WHIM connection: a closer look}
\label{subsec:halo_WHIM_connection}

The analysis of the surface brightness maps and the halo-emitter cross-correlation function shows that a significant fraction of the line emitting WHIM, especially the one that could be detected, is located within or nearby dark matter halos.
We can take advantage of large statistical sample of more than 1 million halos in the \IllBig \ simulation to explore the halo-whim connection in more detail and estimate which fraction of the WHIM emission is actually generated within halos of a specified mass.

To do so we have generated a new set of OVII and OVIII surface brightness maps using only WHIM particles that belong to halos of a given mass interval and count the number of pixels brighter than $S_B$.
To decide whether a particle belongs to a halo, and given the flexible definition of the halo boundaries, we have adopted two separate criteria.
The first one is based on the familiar definition of virial radius, $\Rvir$: a WHIM particle is associated to a halo in the simulation if its distance from the halo center is less than the radius at which the mean density is
200 times larger than the cosmic mean.
The second one uses the definition of ``splashback radius'' $R_\mathrm{sp} $, defined as the distance at which the accreted matter particles' orbits reach the first apocenter after turnaround.
In correspondence of this radius the density profile of the halo becomes steeper than isothermal ($\de\log\rho/\de\log r < -4$), providing an effective boundary to the structure.
The values of $\Rvir$ and  $R_\mathrm{sp} $ are not independent and several fitting formulae have been proposed to specify their relation and the dependence on the mass accretion rate.
Common to these expression is the inequality $R_\mathrm{sp} \lesssim 1.9 \ R_\mathrm{200m}$ which holds true even in the most extreme accretion cases. 
Here, to be conservative and for the sake of simplicity we define and effective splashback radius
$R_\mathrm{sp} \equiv 2\times \Rvir$ and consider all WHIM particles within this distance from the halo centers.

\begin{figure}
	\includegraphics[width=\columnwidth]{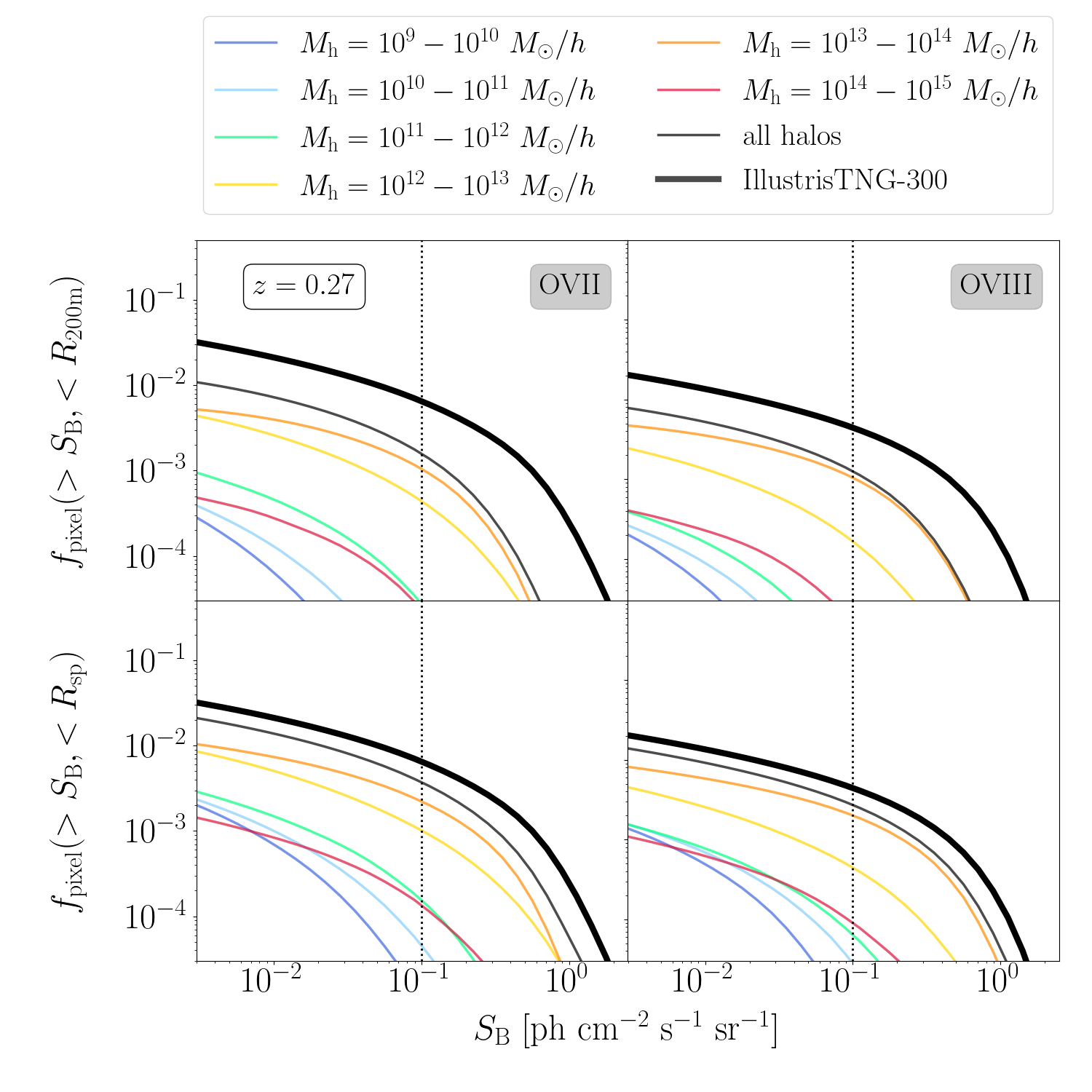}
    \caption{\textit{Top:} Surface brightness $\logNS$ in the \IllBig \ simulation.
    The thick black solid line shows the $\logNS$ of WHIM particles in the \IllBig \ slice.
    The vertical grey shaded area is the $\SB$ region not observable by Athena.
    Thinner lines display the X-ray emission coming from WHIM particles found inside halos (within $\Rvir$) of different mass decades.
    \newline
    \textit{Bottom:} Same as top, but thinner lines display the emission from WHIM particles inside $\Rsp$.}
    \label{fig:CDF_I300_halo_mass}
\end{figure}

In Fig.~\ref{fig:CDF_I300_halo_mass} we show the WHIM line emitter counts breakdown. The thick black curves in all the panels represent the usual $\logNS$ counts of the OVII (left panels) and OVIII (right) lines in the surface brightness map obtained from the \IllBig \ snapshot at $z=0.27$. Only WHIM particles have been considered.
The thin black lines show the $\logNS$ counts obtained when considering only WHIM particles within $\Rvir$ (left panels) and  within $R_\mathrm{sp} $ (right).
Clearly, only a minor, though sizable, fraction of line counts are fully (or mainly) contributed by gas within a halo. The majority of the counts are contributed by multiple structures along the line if sight, not necessarily associated to halos.
This fraction increases up above $\sim 50 \% $ when considering a larger region of radius $R_\mathrm{sp} $, but does not become entirely dominant.

The curves with different colors show the contribution from gas within halos in a specific mass range, indicated in the label.
Interestingly, the magnitude of the contribution increases with the mass of the halo. It is largest for halos in the mass range  $10^{12}-10^{14} \ M_\odot/h$ and then suddenly drops.
This is not surprising since the gravitationally bound mass in OVII and OVIII is remarkably constant for halo masses larger than $10^{12} \ M_\odot/h$  and decreases for lower masses \citep{Nelson+18}. The small contribution from very massive halos juste reflects the paucity of these rare structures.

\begin{figure}
	\includegraphics[width=\columnwidth]{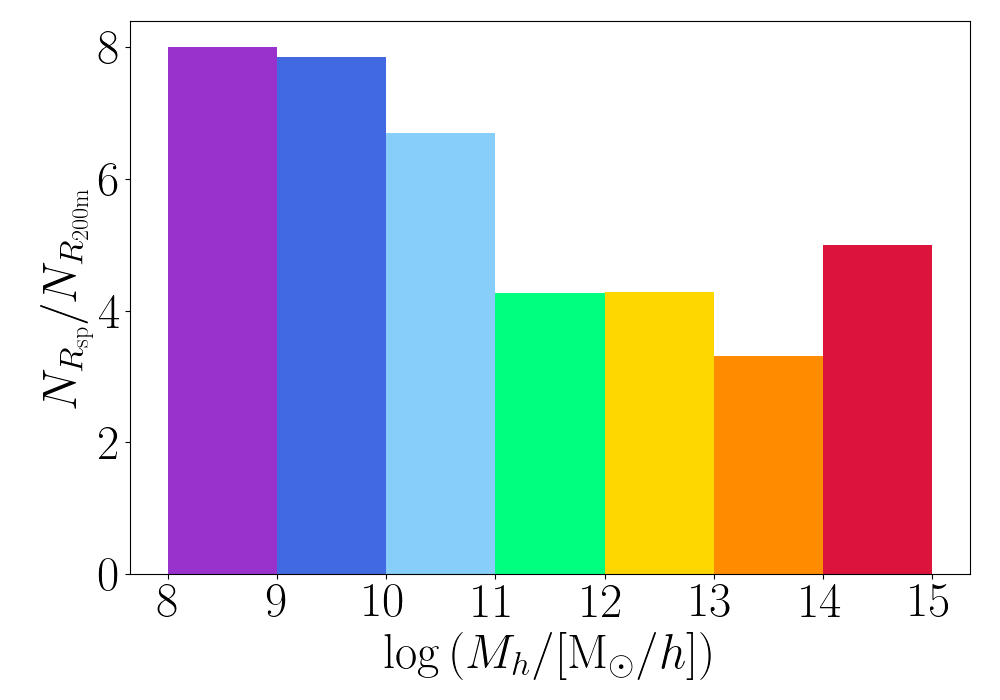}
    \caption{Histogram representing the ratio between number of WHIM particles within 
    $R_\mathrm{sp} $ ($N_{R_\mathrm{sp}}$) and that of the particles within $\Rvir$ ($N_{\Rvir}$) estimated for halos in different mass ranges, indicated in the X-axis. The color code is the same as in Fig.~\ref{fig:CDF_I300_halo_mass}.}
    \label{fig:histogram_hmass}
\end{figure}

To further inspect the WHIM properties in the vicinity of dark matter halos we compare the counts of WHIM particles at $R_\mathrm{sp} $ with the counts at $\Rvir$, effectively sampling the WHIM particles density profile in the outer region of the halos.
The result is shown in Fig.~\ref{fig:histogram_hmass} in which we plot the ratio of the counts as a function of the halo mass. The ratio is not constant. It decreases from $\sim 8$ to a value of about 4, featuring a sudden drop at  $10^{11}\ M_\odot/h$.
The value $\sim 4$ measured for large halos is what one expects for an isolated halo following an isothermal density profile in the interval $[\Rvir, R_\mathrm{sp}]$. That is to say, this value indicates that the WHIM particles that contribute to the line emission belong to the halo.
On the contrary, the larger ratio $\sim 8$ associated to less massive halos, indicates a density of WHIM particles higher than that expected for a isolated halo and that a significant fraction of the mass in the outer regions is probably associated to the large scale structures in which the halo is embedded.

\begin{figure}
	\includegraphics[width=\columnwidth]{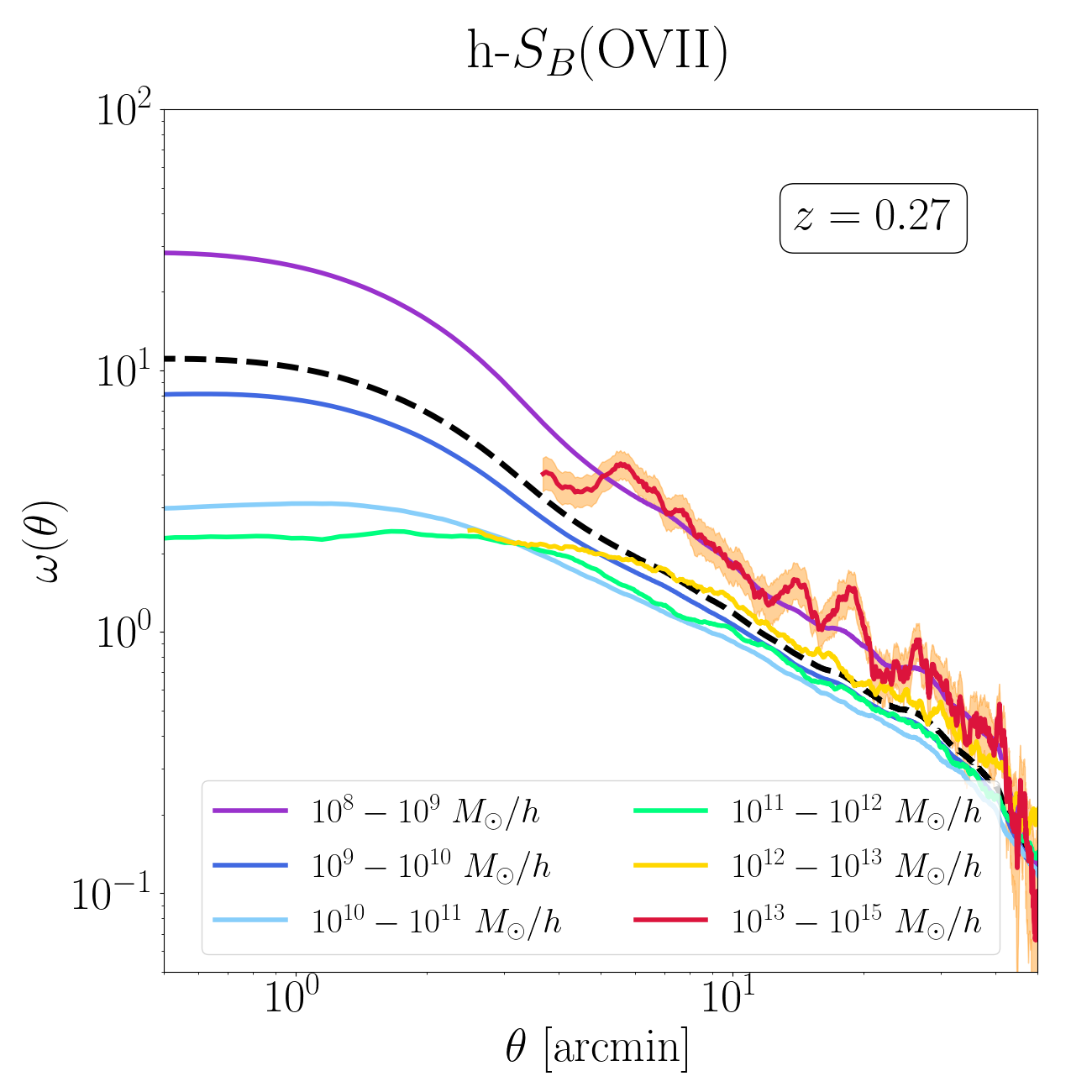}
    \caption{Angular cross-correlation between OVII emitters and the centers of dark matter halos of different masses in the \IllBig \ snapshot at $z=0.27$. 
    Different colors identify different mass ranges, indicated in the label. The color code is the same as in Fig.~\ref{fig:CDF_I300_halo_mass}. 
    Halos with mass in the mass range $[10^{13},10^{15}]$ $M_\odot/h$ have been grouped together because of their limited number. The orange strip around the ref curve shows the scatter estimated from 196 jackknife resampling the same halo-emitter datasets. The thick dashed line is represents the corss-correlation of all OVII emitters with all the halos in the sample.}
    \label{fig:angular_2PCFs_IllustrisTNG300_halo_mass}
\end{figure}

Fig.~\ref{fig:angular_2PCFs_IllustrisTNG300_halo_mass} clarifies the issue further.
It shows the angular OVII-halo angular cross-correlation function for halo of different masses, i.e. the angular surface brightness profile of the WHIM OVIII emission around halos in the same mass ranges as in Fig.~\ref{fig:histogram_hmass}. 
The different brightness profiles have similar shapes: a central core-like structure followed by a power-law decrease.
Less massive halos have shallower profile, whereas the central region of the the more massive ones could not be probed because of the large halo-OVII emitter mean separation.
The point is that the $[\Rvir, R_\mathrm{sp}]$ interval corresponds, for small halos, to 
sampling the core-like region where the OVII surface brightness profile is shallower.
On the contrary, for larger halos, the same range of physical separations probes the outer region of the surface brightness profile, where the slope is steeper. That is to say, a significant fraction of the WHIM gas is associated to the the large scale structures (external halos, filaments) rather than to the halo itself.
We conclude that the  bright WHIM line emission associated to prominent structures in the surface brightness maps is mostly associated to the halo mass. On the contrary, emission associated to smaller structures is significantly contributed by structures of larger scale. 


\section{$\logNS$ forecast on a lightcone}
\label{subsec:lightcone}

\begin{figure*}
	\includegraphics[width=\textwidth]{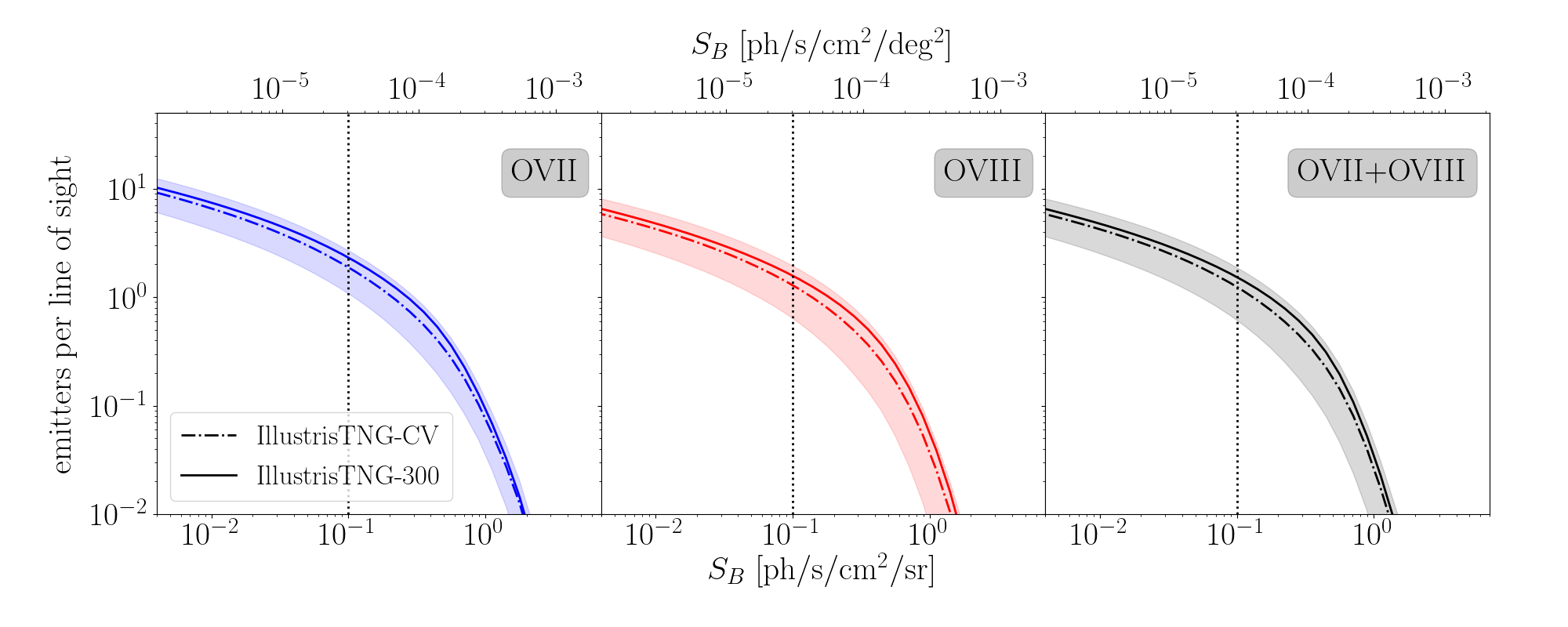}
    \caption{Number of
    WHIM emitters in OVII (left), OVIII (center) and OVII+OVIII (right) per line of sight as a function of surface brightness in a lightcone to $z=0.54$.
    Predictions from the \IllBig (continuos curves) and \IllCV (dashed) simulations
    are shown surrounded by a shaded region representing the {it rms} scatter among the 
    27 \IllCV realizations. 
    The vertical black dotted line identifies the expected 3-$\sigma$ detection threshold of a 100 ks observation with Athena.}
    \label{fig:emitters_per_los}
\end{figure*}

So far we have simulated surface brightness maps at well defined redshifts, obtained from a few selected snapshots of the CAMELS simulations.
Real maps will, instead, account for line emission from all redshift along the lightcone. 
Therefore, to obtain more realistic predictions on the possibility of studying the WHIM in the X-ray band and to compare our results with those in the literature, we decided to simulate the surface brightness maps of the OVII and the OVIII lines from all the gas particles along the line of sight to $z=0.54$. To do so we have rearranged the various simulated outputs into a lightcone using the approximated procedure described below. We built two types of lightcones: those obtained from 9 snapshots of the \IllCV simulations set and the one obtained from the 32 snapshots of the \IllBig \ simulation. The procedure adopted is the same in both cases, only the thickness of the snapshots slices piled up to build the cone is different. 
Also, since our procedure relies on the line counts performed on the individual snapshots, for which we have assumed the distant observer hypothesis, we cannot use the very nearby $z=0$ snapshot and build the lightcone using only the simulation outputs in the range $0.05\leq z \leq 0.54$.
The procedure to simulate the brightness maps and estimate the line counts is as follows:
\begin{itemize}
    \item We define the footprint of the lightcone. This is a squared area with an opening angle  $\arctan\left(\frac{L/2}{\chi(z=0.05)}\right)$, where $\chi(z=0.05)$ is the comoving distant to the redshift to the nearest snapshot in the cone.
    \item Given the available snapshots, we define the redshift interval $[z_{i-1},z_i]$,    where $z_i$ is the redshift of a snapshot, except for $z_i=0.05$ for which the corresponing simulation snapshhot is the one at $d=0$.

    \item Having defined the area and the thickness of all non-overlapping slices we estimate their individual volume $V_i$.
    \item We rescale the oxygen line counts already performed on each snapshot $N(>\SB, z_i)$ to the volume of each redshift slice $N_i(>\SB,z_i)\rightarrow N(>\SB, z_i)\frac{V_i}{L^2\Delta L}$, where $\Delta L$ is the thickness of the simulation slice at each redshift (see Table \ref{tab:snapshot_grid}).
    \item We repeat the procedure for all the slices, sum over, divide by the number of pixels in the footprint and obtain the number of OVII and OVIII lines brighter than $S_B$ per line of sight shown in Fig. \ref{fig:emitters_per_los}.
\end{itemize}

The thick line in the plots represents the $\logNS$ of the OVII (left), OVIII (center) and joint OVII+OVIII (right) line counts from the lightcone extracted from the \IllBig \ simulation.
The size of the simulation box is about 6 times smaller than the comoving distance to $z=0.05$, which inevitably implies some replication of the same structures in the lightcone.
Since we are interested in the line counts only, we made no attempt to rearrange the orientation of the slices across the cone.
On the other hand, replicating the same structures artificially removes cosmic variance, to appreciate which we use the 27 lightcones obtained from the 
\IllCV simulations.
The resulting average counts predictions are represented by the dashed curves in the panels and the surrounding shaded areas the magnitude of the cosmic variance.
It is reassuring that both $\logNS$ curves, from the  \IllCV and the \IllBig \ lightcones are consistent with each other. Both exhibit a characteristics double power-law shape with a knee at $\SB\sim 0.3 \ \unitsSB$ beyond which the line counts become more rare.
With the X-IFU 3-$\sigma$ sensitivity threshold set at $\SB\sim 0.1 \ \unitsSB$ for a 100 ks observation with Athena (vertical gray band) the number of expected detections per pixel is in the range [1,3] for OVII and [1,2] for OVIII (and OVII+OVIII).
Multiplying by the number of pixels in the X-IFU field of view 
(with a diameter of 5 arcmin) we then expect $[3600,10800]$ OVII detections 
and $[3600,7200]$ OVII (and OVII+OVIII) detections in total.

\begin{figure}
	\includegraphics[width=\columnwidth]{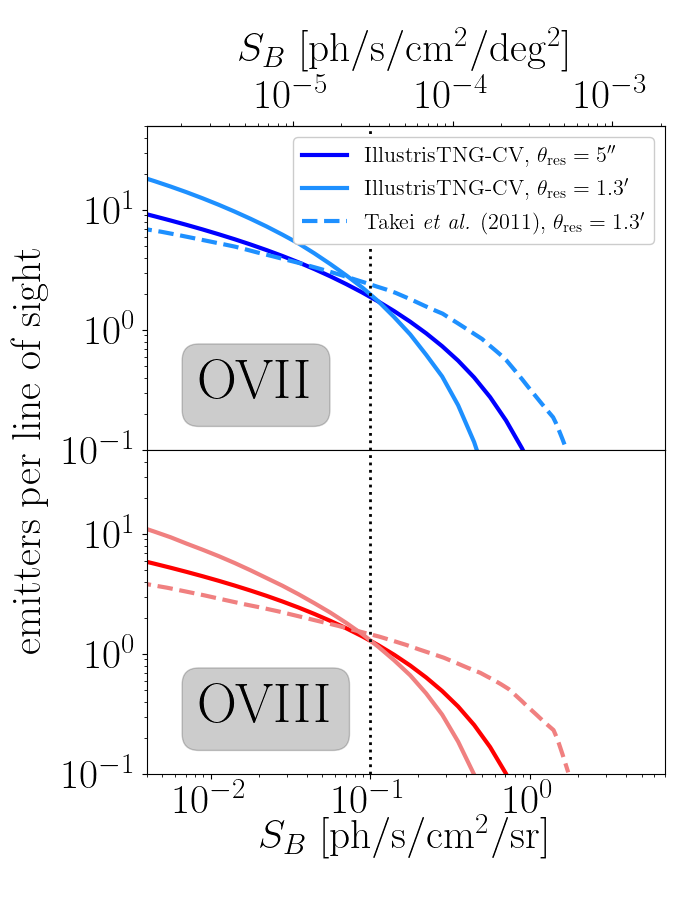}
    \caption{$\logNS$ curves for OVII (top panel) and OVIII (bottom) representing the average line counts in pixels of angular size 5 arcsec (dark curves) and 1.3 arcmin (light curves) in the \IllCV realizations. The dashed curve is plotted for reference and represents the analogous $\logNS$ counts of \protect\cite{WHIM-Takei+11} as shown in Fig. 2 of their paper.}
    \label{fig:emitters_per_los_literature}
\end{figure}

These counts, however, should not be regarded as independent since in most cases line emission in nearby pixels is produced by a single physical structure.
Previous works have shown that WHIM emitters have a typical angular size of a few arcminutes \citep{Ursinogaleazzi06,galeazzi09,Ursino+11,WHIM-Takei+11}, also confirmed by our angular auto-correlation analysis. For this reasons the first proposals to detect the WHIM in emission with X-ray integral field spectroscopy (like  e.g. {\it EDGE} \citep{edge09} or {\it Xenia} \citep{xenia10}) were designed to have a matching angular resolution and theoretical forecast where provided accordingly.
As a result, to compare our $\logNS$  with those of \cite{WHIM-Takei+11} we had to rescale our predictions to the angular resolution of 1.3 arcmin of their simulated maps.
The results are shown in  Fig. \ref{fig:emitters_per_los_literature} where, along with our original counts in 5 arcsec pixels (continuous dark curves) we show the counts in the larger 1.3 arcmin pixels (continuous light curves), both superimposed to the
$\logNS$  reported from model B2 of \cite{WHIM-Takei+11} also estimated from a map with 
1.3 arcmin resolution.
All curves account for the number of OVII and OVIII emission lines generated by WHIM gas in a lightcone out to $z=0.54$.

Decreasing the angular resolution increases the faint line counts and decreases the bright ones. This is not surprising. The integrated emission over a 1.3 arcmin regions will generally be composed by a number of brighter and fainter pixels that contribute to correspondingly bright and faint counts in the 5 arcsec resolution $\logNS$ curves.
The comparison with the result of \cite{WHIM-Takei+11} shows that the two predictions are quite similar for the faint counts whereas in the bright end we predict systematically less counts, regardless of whether we use of \IllCV or SIMBA-CV.
This mismatch is not surprising. \cite{Roncarelli+12} also noticed that the bright counts of  \cite{WHIM-Takei+11} B2 models were systematically larger than theirs.
The likely explanation is related to the details of the WHIM model. Part of the reason is related to the simulation technique: the model of \cite{WHIM-Takei+11} relies on the hydrodynamic simulations by \cite{WHIM_simulations_GADGET2-Borgani+04} which uses
the GADGET-2 \citep{gadget-3_2005} code, as opposed to the IllustrisTNG simulation  that have been performed with the AREPO code. However, this fact alone does not explain the mismatch, since as we already mentioned using the counts from the \SIMCV maps instead of the \IllCV ones would not erase the difference.
The main effect is likely related to the treatment of the gas metallicity.
\cite{WHIM-Takei+11} did not assign gas metallicity in a self consistent fashion.
They did so in the post-processing phase using the phenomenological density-metallicity relations derived from the \cite{Cosmic_chemical_evolution-Cen+99} hydrodynamic simulation. It has been shown \citep{Roncarelli+12} that the large scatter of that relation, assumed to be Gaussian, artificially enhanced the gas metallicity in low density environment, making the WHIM emission brighter.

One remarkable feature, though, is the fortuitous crossing of the various $\logNS$ curves in correspondence to a surface brightness value similar to the Athena X-IFU detection threshold, which implies that the number of expected detections per line of sight is similar to that predicted by \cite{WHIM-Takei+11} and insensitive to the angular resolution of the map.

\section{Discussion and conclusions}
\label{sec:conclusions}

There is now a general consensus that the bulk of missing baryons are to be found in the form of a highly ionized warm hot gas in the intergalactic medium that could be effectively probed by means of X-ray spectroscopy. 
The recent claim of missing baryons detection from the absorption lines of X-ray and UV spectra of distant quasars 
\citep{WHIM_detection-Nicastro+18,WHIM_detection-Kovacs+19} are regarded as genuine and confirm that the proposed solution to the missing baryon problem is correct.
However, a few detections along two sightlines are largely insufficient  to address all the open issues regarding the spatial distribution of baryons in the late Universe, their ionization state, metal enrichment and thermal history.
Only next generation instruments will be able to combine absorption studies capable to probe missing baryons in moderate density environments along selected sightlines and emission studies of the gas in high density regions across a sky patch can tackle the problem.
The possibility of performing  X-ray integral field spectroscopy with proposed space missions like XRISM \citep{Xrism+20} and, especially, Athena \citep{Athena} and LEM will give us such opportunity.

Several works have investigated the feasibility of studying the WHIM with next generation instruments using hydrodynamic simulations. And most of them have focused on the possibility to do so in absorption.
In this work we took advantage of the availability of a very large number of hydrodynamic simulations from 
the CAMELS project \citep{CAMELS_simulations, CAMELS_simulations_data_release}, 
to systematically investigate the properties of the WHIM in emission over a wide range of scenarios characterized 
by different choices of cosmological parameters as well as stellar and AGN energy feedback models, and to assess the significance of their impact with respect to the expected cosmic variance. Moreover, the possibility of comparing results obtained using identical conditions but different simulation techniques provides us with the unique opportunity to assess the robustness of the model predictions to the choice of the numerical tools and allows us to estimate the theoretical systematic errors.

One important aspect of the CAMELS simulations is the size of the computational box, 25 comoving  Mpc/$h$, which is much smaller than the size of a typical cosmological simulation but more than adequate to investigate the possibility of studying the WHIM in emission in future observations considering the limited field of view 
($\sim 5$ arcmin$^2$) according to the the current X-IFU design.

The main results of our analysis are as follows:
\begin{itemize}
    \item All the OVII and OVIII line surface brightness maps share the same qualitative features. The spatial distribution of the WHIM emission traces that of the dark matter halos. Only the gas concentrated around the halos is potentially detectable. The evolution of the WHIM gas from $z\simeq 0.5$ to the present epoch is rather mild. None of these aspects is new. \cite{WHIM-Takei+11} have been the first to point out that emission studies could not detect the WHIM in the low density, large scale filaments  and, instead, would preferentially probe the gas near the halos at the nodes of the cosmic web. Similarly, \cite{martizzi19} have shown, using the same IllustrisTNG simulations considered in this work, that the mass fraction of WHIM in knots and filaments evolves little between $z=0$ and $z=1$. What is new here is the fact that these features are robust to the choice of the cosmological parameters, namely $\sigma_8$ and $\Omm$, to the parameters that regulate the energy injection in the IGM from star formation and AGN activities and to the numerical method adopted to simulate the hydrodynamic evolution of the system.
    
    \item Line counts as a function of their surface brightness $\SB$ are a useful statistics that can be determined from all the surface brightness maps at different redshifts. They can be extracted from the data cube generated by an instrument like X-IFU. We estimated these counts from the simulated maps and found a significant dependence on the type of simulation considered. The WHIM line counts in the \IllCV realizations are systematically larger than in their \SIMCV analog. The difference is significant, i.e. it is comparable to the expected cosmic variance in a field of view as large as the simulation box.  
    It is also of the same order as the counts variations induced by changing the parameters of the models, including those that regulate the energy feedback.
    This result highlights the importance of including, in the CAMELS suite, numerical experiments performed with different techniques. Precisely to highlight possible tensions that, if not solved, need to be included in the model error budget.  
    
    By contrast, line number counts predictions are quite robust to varying the values of $\sigma_8$ and $\Omm$ parameters within their (generous) allowed intervals. Instead, they are sensitive to the details of the energy release process: increasing the amount of energy injected through star formation and supernovae explosion significantly increases the number of bright line counts (and decreases the faint one). What is perhaps more surprising is that the energy release from AGN activity has the opposite effect. This interplay between the two phenomena and its redshift dependence can be key to appreciate which one is more relevant and when.
    Furthermore, our results are converged as a function of cosmological volume probed, as our comparison between \IllCV and \IllBig \ testifies.
    
    \item The spatial correlation properties of the OVII and OVIII lines and their distribution around dark matter halos have been inferred from the measured 2-point auto and cross-correlation function between emitters and halos. Although some mismatch exists at small separations, the similarity between the \IllCV and the 
    \SIMCV predictions is very reassuring and suggests that whatever the reason for the line count mismatch may be, it originates from the central regions of the dense structures, where the number of close pairs is larger. Since the clustering analysis of the WHIM lines uses the redshift as a proxy to the emitter distance, it suffers from redshift space distortions. Therefore, one can use measure the quadrupole moment of the 2-point cross-correlation function to infer the dynamics of the gas relative to halos. A further advantage is that the quadrupole is not sensitive to the integral constraint that, instead, forces the monopole amplitude to cross zero at relatively small separations, making it difficult to trace the distribution of the gas far away from the halos.
    The results show that gas dynamics near the center of the halos is characterized by incoherent motions (positive quadrupole) and it is completely different from that of the outer regions, where coherent infall motions characterize the accretion of the gas onto the knots and the dense filaments of the web.
    
    \item The energy resolution of the spectrometer hampers our ability to probe the spatial distribution of the WHIM. However, it permits to perform a tomographic analysis of its clustering properties based on the angular 2-point correlation function of the surface brightness maps. We have analysed the angular correlation properties of the OVII and OVIII emitters in our simulated maps and found that the correlation signal is potentially detectable using next generation instruments. In fact, the signal-to-noise can be amplified by cross-correlating the line surface brightness with the angular position of extragalactic objects located at or near the center of galactic halos, suggesting that a strategy to detect the WHIM alternative to the stacking technique of \cite{tanimura2020}, is that of cross-correlating the X-ray maps observed in some specific (and conveniently narrow) energy band with the angular position of galaxies in a shell conveniently extracted from a spectroscopic galaxy sample.
    
    The measured angular auto- and cross- 2-point correlations are sensitive to $\sigma_8$ and $\Omm$ in the sense that their increase enhances the angular correlation amplitude at small angular separations. This is expected since increasing the amount of matter and the amplitude of its density fluctuations accelerates the structure formation process and the accretion rate of the WHIM gas. What is perhaps more surprising is the robustness of the WHIM angular correlation properties to the amount of energy released in the medium by star formation processes and AGN activities. Unlike the case of line counts, changing any of the four parameters that regulate the amount of energy feedback in our model has no significant impact on the spatial correlation properties of the line emitting WHIM.
    We interpret this insensitivity as an evidence that the properties of the WHIM, and more specifically its spatial distribution, are largely determined by the gas accretion and subsequent formation of the shock fronts during the nonlinear structure formation, a process that is captured by the the hydrodynamic and thermal aspects of the simulation and little sensitive to the details of the energy feedback processes.
    
    \item To check the sensitivity of our results to the (small) volume of the simulations we compared our results with those obtained from a larger simulation, \IllBig, and found no significant differences. In doing so, we have exploited the better statistical sample of dark matter halos in the extracted from the larger box and investigate in detail the halo-WHIM connection. We found that a significant, though not dominant, fraction of the WHIM emission comes from within the virial radius $\Rvir$ or from the so-called splashback radius, that we place at a distance of $2 \times \Rvir $. 
    Most of this emission is associated to halos in the mass range $10^{12}-10^{14} \ M_\odot/h$ and seems to come from the halo itself since the number density of the WHIM particles seem to follow the mass density profile of the halo. Smaller halos provide a minor contribution to the line emission. However, not all the material responsible for such emission seem to be associated to the halo and, quite possibly, traces the mass distribution within larger structures in which halos are embedded, 
    
    \item To compare our $\logNS$ to those of \cite{WHIM-Takei+11} we have computed the line counts on a lightcone to the same $z=0.54$ and matching angular resolution of 1.3 arcmin. This resolution was originally chosen to match the typical angular size of the WHIM emitters and maximize the signal-to-noise of the line counts. Decreasing the angular resolution from that of the X-IFU instrument (5 arcsec) to 1.3 arcmin  decreases the the number of bright lines per line of sight and increases the faint counts, as expected. What is remarkable, and fortuitous, is the fact that the number of lines above $0.1 \ \mathrm{photons \ cm^{-2} \ s^{-1} \ sr^{-1}}$, which represents a 3-$\sigma$ detection threshold for a 100 ks observations with Athena X-IFU, does not change with the angular resolution of the map and is similar to that predicted by \cite{WHIM-Takei+11}. On the other hand, the number of counts brighter than
    the threshold in \cite{WHIM-Takei+11} is significantly larger than in our case, irrespective of the resolution considered or the type of simulation, or the energy feedback model adopted.
    \cite{Roncarelli+12} have noticed a similar, though qualitative, mismatch with their bright counts and argued that the optimistic predictions of the \cite{WHIM-Takei+11} model originated from the treatment of the gas metallicity in the simulation that, unlike our case, was not determined during the run but assigned in the post-processing phase.
     
    The high resolution of the X-IFU maps is not designed to optimize the probability of the WHIM detection since, as we have pointed out, it does not match the typical size of the WHIM emitter. And it comes at the price of a small field of view, effectively hampering the possibility to trace the distribution of the WHIM over cosmological volumes. On the other hand, it allows one to spot and remove ``bad'' pixels in which the signal is contributed by bright, point-like sources like the AGNs. Not just that: if the goal is to investigate the WHIM properties, having a better resolution will help separating the pixels in which the emission is dominated by the WHIM component from those dominated by non-WHIM gas emission according, of course, to some model dependent criteria. Separating these components is important, for example, if one wants to extract the angular correlation function of the WHIM which is subdominant with respect to the total emission of the whole gas \citep{Ursino+11}. If flown, LEM will have similar angular and energy resolution to the Athena X-IFU, but across a much wider field of view of 32~arcmin, enabling large sections of the WHIM to be studied in single pointings.

\end{itemize}

In this work we did not attempt to provide realistic forecasts on the ability to detect and study the WHIM in emission with some specific instrument, although we often referred to the case of a ``typical'' 100 ks observation with Athena X-IFU. We feel, however, that it will be important to do so in the future, once next generation X-ray instruments will be approved and their technical design will be frozen. It will be important to include a realistic treatment of the foreground emission, especially from the Galaxy, and re-derive a more realistic detection threshold for all potentially observable X-ray lines, not just OVII and OVIII.
Nevertheless, we believe that the main results of this work that we have summarized above are robust and that our main conclusions will not be significantly modified using more realistic datasets.
What we plan to do next is to improve the statistical analysis of the surface brightness maps to increase the statistical significance of the WHIM detection and improve our ability to study its physics. One possibility would be to perform a joint statistical analysis that includes both line counts and angular correlation and their redshift dependence. 
Additionally, one could combine the line emission analysis with that of other WHIM probes like 
the Sunayev Zel'dovich effect \citep{CGM_CAMELS-Moser+22}.
Better still, one could fully exploit the power of the CAMELS simulation suite and apply machine learning techniques to the surface brightness (and possibly Sunayev Zel'dovich) maps
to perform field-level likelihood-free inference.

\section*{Acknowledgements}
This work is supported  by ``Attivit\`a di Studio per la comunit\`a scientifica di Astrofisica delle Alte Energie e Fisica Astroparticellare  Accordo Attuativo ASI-INAF n. 2017-14-H.0'' and by the INFN project ``InDark''.
GP and EB  are also supported by MIUR/PRIN 2017 ``From Darklight to Dark Matter: understanding the galaxy-matter connection to measure the Universe''. The CAMELS project is supported by NSF grants AST-2108944, AST-2108678, and AST-21080784. We acknowledge Shy Genel for timely technical help.
We thank Veronica Biffi, Daisuke Nagai, Benjamin Oppenheimer, Benjamin Wandelt, Erwin Lau, Nicholas Battaglia, David Spergel, Pierluigi Monaco and Dylan Nelson for useful discussions.



\bibliographystyle{mnras}
\bibliography{bibliography}

\label{lastpage}
\end{document}